\documentclass[useAMS,usenatbib,referee]{biom}
\usepackage{tikz,pgf}
\usepackage[figuresright]{rotating}
\usepackage{graphicx}
\usepackage{endfloat}
\usepackage{amssymb}
\usepackage{amsmath}
\usepackage{graphicx}
\usepackage{verbatim}
\usepackage{subcaption}
\usepackage{float}
\usepackage{placeins}
\usepackage{url}
\usepackage{physics}
\usepackage{booktabs}
\usepackage{color}
\usepackage{thmtools}

\DeclareMathOperator{\pa}{pa}
\DeclareMathOperator{\nd}{nd}
\DeclareMathOperator{\pre}{pre}

\def\ci{\perp\!\!\!\perp}

\newtheorem{thm}{Theorem}
\newtheorem{lem}[thm]{Lemma}

\newif\showComments
\ifdefined\showComments
	\newcommand{\mycomment}[3]{\textcolor{#1}{[\textbf{\textsc{#2}}: \textit{#3}]}}
\else
	\newcommand{\mycomment}[3]{}
\fi

\usepackage{framed} 

\newcommand{\optionrule}{\noindent\rule{1.0\textwidth}{0.75pt}}

\newenvironment{aside}{%
  
  \MakeFramed {\advance\hsize-\width \small}\optionrule}
{\newline\optionrule\endMakeFramed}

\begin{document}

\title[Markov-Restricted Analysis of Randomized Trials with Non-Monotone Missing Binary Outcomes]{Markov-Restricted Analysis of Randomized Trials with \\ Non-Monotone Missing Binary Outcomes}

\author{Jaron J. R. Lee$^1$, Agatha S. Mallett$^2$,   Ilya Shpitser$^1$ \\ {\bf Aimee Campbell$^3$,  
Edward Nunes$^3$,   Daniel O. Scharfstein$^{4,*}$\email{daniel.scharfstein@hsc.utah.edu}}  \\
$^1$Department of Computer Science, Johns Hopkins University, Baltimore, MD U.S.A. \\
$^2$Department of Computer Science, University of Utah Kahlert School of Computing, Salt Lake City, UT, U.S.A \\
$^3$Department of Psychiatry, Columbia University Irving Medical Center and \\ New York State Psychiatric Institute, New York, NY, U.S.A \\
$^4$Department of Population Health Sciences, University of Utah School of Medicine, Salt Lake City, UT, U.S.A.}

\begin{abstract}
\cite{scharfstein2021global} developed a sensitivity analysis model for analyzing randomized trials with repeatedly measured binary outcomes that are subject to nonmonotone missingness. Their approach becomes computationally intractable when the number of measurements is large (e.g., greater than 15).  In this paper, we repair this problem by introducing $m$th-order Markovian restrictions.  We establish identification results for the joint distribution of the binary outcomes by representing the model as a directed acyclic graph (DAG).  We develop a novel estimation strategy for a smooth functional of the joint distribution. We illustrate our methodology in the context of a randomized trial designed to evaluate a web-delivered psychosocial intervention to reduce substance use, assessed by evaluating abstinence twice weekly for 12 weeks, among patients entering outpatient addiction treatment. \\ 
\end{abstract}


\maketitle

\newpage

\section{Introduction}
\label{sec1}

Non-monotone missing data are a common occurrence in randomized trials in which study participants are scheduled to be assessed at fixed time points after randomization.  One of the most common assumptions used to identify treatment effects in such studies is the missing at random (MAR) assumption.  While MAR has been considered a reasonable benchmark assumption for longitudinal studies that have monotone missing data patterns, \citet{robins1997non}, \citet{Vans:Rotn:Robi:esti:2007} and \citet{little2014statistical} have argued that MAR is implausible for studies that have non-monotone missing data patterns. In large part, this is because MAR ignores the temporal ordering of the variables. 

A variety of missing not at random (MNAR) assumptions have been proposed for analyzing studies with non-monotone missing data (see, for example, \cite{little1993pattern,robins1997non,Vans:Rotn:Robi:esti:2007,zhou2010block,shao2012imputation,sadinle2016itemwise,shpitser2016consistent,tchetgen2016discrete,sadinle2017sequential,linero2018bayesian, scharfstein2021global}).  In this paper, we focus on the temporally ordered version of the MNAR assumption introduced by \citet{robins1997non}.  This assumption states that the probability of missing a given assessment depends only on the outcomes (observed or not) prior to the assessment and the observed data after the assessment. \citet{scharfstein2021global} introduced a set of identifying assumptions, indexed by sensitivity analysis parameters, that is anchored around the \citet{robins1997non} assumption.  In the context of binary outcomes, \citet{scharfstein2021global} developed an inferential strategy based on estimating the distribution of the observed data using random forests and functionals of the full data distribution using the plug-in principle; they established $\sqrt{n}$-asymptotic theory. Their use of random forests was a flexible way of addressing the curse of dimensionality \citep{robins1997toward}. 
Unfortunately, this approach will not work when the the number of post-baseline assessment times, $K$, is ``large".  This is because their algorithm requires storage and operation on a $3^K$ vector of probabilities, which we have found to be computationally intractable when $K>15$. 


In this paper, we address this problem by introducing $m$th-order Markovian-type conditional independence restrictions. We represent these restrictions using directed acyclic graph (DAGs).  The strategy we adopt for showing identification of target parameters in our restricted model yields identifying functionals expressible in terms of conditional distributions 
derived from the observed data law. These conditional distributions have size that is bounded by a constant that depends on $m$, rather than $K$.  Our approach is instead linear in $K$ and exponential in a smaller constant $m$. 



The outline of the paper is as follows. In Section 2, we introduce notation.  In Section 3, we discuss the Markov-restricted version of the class of models introduced by \citet{scharfstein2021global}.  Each model in the class can be represented as a DAG.  In Section 3, we prove, using the properties of DAGs, that the full data law is identified.  In Section 4, we discuss estimation.
Section 5 presents a re-analysis of CTN-0044, a randomized trial designed to evaluate 
a web-delivered psychosocial intervention to reduce substance use
among patients entering outpatient addiction treatment \citep{campbell2014internet}.  In this study, substance use was scheduled to be collected twice weekly for 12 weeks (i.e., $K=24$).  Section 6 presents the results of a realistic simulation study.  Section 7 is devoted to a discussion.

\section{Notation}

  Let $Y_k$ denote the binary outcome (possibly unobserved) at assessment $k$ ($k=1,\ldots,K$).  Let $R_k$ be the binary indicator that $Y_k$ is observed.    Let $Y^{obs}_k = Y_k$ if $R_k=1$ and $Y_k^{obs} =~?$ if $R_k=0$.  Let $O_k$ be the observed data at assessment $k$; it can be represented by $(R_k, Y^{obs}_k)$.
 We note that $Y_k^{obs}$ is a deterministic function of $R_k$ and $Y_k$.

For any given vector $\overline{z}= (z_1 , z_2 , . . . , z_K)$, define 
$\overline{z}_k = (z_1 , . . . , z_{k-1})$, 
$\underline{z}_k = (z_{k+1} , . . . , z_K)$, 
$\overline{z}^m_k = (z_{\max(1,k-m)} , . . . , z_{k-1})$, 
and
$\underline{z}^m_k = (z_{k+1} , . . . , z_{\min(k+m,K)})$, where $m$ denotes the order of the Markovian restiction introduced in the next section. 
With this notation, note that $\overline{z}_k^m = \underline{z}_{k-m-1}^m$. 
There will be instances in which we refer to a vector in which the lower index is larger than the upper index; in this case, the vector will be taken to be the null vector.

We denote the observed data vector for an individual 
as $\overline{O}$.  The goal is to use $n$ i.i.d. copies of $\overline{O}$ to draw inference about a given functional of the distribution of $\overline{Y}$.

In what follows, we will use the notation $f(\cdot)$ and $f(\cdot | \cdot)$ when referring to the marginal and conditional distribution of a random variable/vector, respectively.  In contrast, we will use the notation $P(A)$ and $P(A|\cdot)$ when referring to the marginal and conditional probability of an event $A$, respectively. Say that $Z$ is a discrete random variable/vector; then $f(Z)$ will refer to the distribution of $Z$ and $P(Z=z)$ will refer to the probability that $Z$ takes on the specific value $z$. 

\section{Markov-Restricted Sensitivity Analysis Model}



We express restrictions in our model using the chain rule factorization of the full data distribution using the following total ordering on variables:
\begin{align}
Y_1 \prec Y_2 \prec \ldots Y_K \prec R_K \prec Y_K^{obs} \prec R_{K-1} \prec Y_{K-1}^{obs} \prec \ldots R_1 \prec Y_1^{obs}. 
\label{eqn:order}
\end{align}

Our $m$th-order Markovian model posits the following restrictions on this factorization:  
\begin{equation}
\label{eqn:permutation-model0a}
f(Y_k \; | \;  \overline{Y}_{k}) =  f(Y_k \; |\;  \overline{Y}^m_{k}) 
\end{equation}
\begin{equation}
\label{eqn:permutation-model0b}
P(R_k=0 \; | \;  \overline{Y}, \underline{O}_k) = P(R_k=0 \; | \;  \overline{Y}^{m+1}_{k+1}, \underline{O}^m_k) 
\end{equation}
and
\begin{equation}
\mbox{logit}\left\{ P(R_k=0 |  \overline{Y}^{m+1}_{k+1}, \underline{O}^m_k) \right\}  = h_k(\overline{Y}^m_{k}, \underline{O}^m_k;\alpha_k) + \alpha_k Y_k,
\label{eqn:permutation-model2}
\end{equation}
where 
\[
h_k(\overline{Y}^m_{k}, \underline{O}^m_k;\alpha_k) = \mbox{logit} \left\{ P(R_k=0 | \overline{Y}^m_{k}, \underline{O}^m_k) \right\} - \log \left\{ c_k(\overline{Y}^m_{k}, \underline{O}^m_k;\alpha_k) \right\}
\]
and 
\begin{equation}c_k(\overline{Y}^m_{k}, \underline{O}^m_k;\alpha_k) = E[ \exp \{ \alpha_k Y_k \} | R_k=1, \overline{Y}^m_{k}, \underline{O}^m_k] \label{eqn:permutation-c-factor}.
\end{equation} 
Importantly, an equivalent representation of the restriction (\ref{eqn:permutation-model2}) may be obtained by Bayes rule as:
\begin{equation}
\label{eqn:permutation-model0c}
f(Y_k | R_k=0, \overline{Y}^m_{k}, \underline{O}^m_k) = \frac{ f(Y_k | R_k=1, \overline{Y}^m_{k}, \underline{O}^m_k) \exp \{ \alpha_k Y_k \} }{ c_k(\overline{Y}^m_{k}, \underline{O}^m_k;\alpha_k)  }
\end{equation}
To enable tractable inference, we consider models where $2m+1 < K$.


Our model is defined by Markov restrictions on factors of the chain rule factorization.  Such models may be represented by directed acyclic graphs (DAGs), where random variables in the model are associated with vertices in a DAG.

\begin{aside}
\noindent {\bf DAG Models:} Given a total ordering $\prec$ on a vector of variables $\overline{V}$, define $\pre_{\prec}(V)$ to be predecessors of $V$ in $\prec$. Let $f(\overline{V})$ denote the distribution of $\overline{V}$.
Define a DAG ${\cal G}$ where for every vertex $V$, the set of parents $\pa_{\cal G}(V)$ of $V$ (vertices in ${\cal G}$ with edges pointing into $V$) is a subset of $\pre_{\prec}(V)$ such that
$V$ is independent of $\pre_{\prec}(V) \setminus \pa_{\cal G}(V)$  given $\pa_{\cal G}(V)$ in $f(\overline{V})$.  
This implies that the chain rule factorization $f(\overline{V}) = \prod_{V \in \overline{V}} f(V \mid \pre_{\prec}(V))$ that uses the order $\prec$ may be reformulated as the DAG factorization $f(\overline{V}) = \prod_{V \in \overline{V}} f(V \mid \pa_{\cal G}(V))$.  The statistical DAG model of ${\cal G}$ consists of any $p(\overline{V})$ that obeys the DAG factorization with respect to ${\cal G}$.
Any such $f(\overline{V})$ obeys the global Markov property, where conditional independencies in $f(\overline{V})$ may be read off from ${\cal G}$ via the d-separation criterion \citep{pearl88probabilistic}.
\end{aside}

Under (\ref{eqn:order}) - (\ref{eqn:permutation-model0c}), the full data 
distributions in our model factorize as:
\begin{equation}
\prod_{k=1}^K f(Y_k | \overline{Y}_k^m) f(R_k | \overline{Y}_{k}^{m}, Y_k, \underline{O}_k^m) {\color{red} f(Y^{obs}_k | R_k, Y_k) }.
\label{eqn:dag-f}
\end{equation}
where the red set of factors are deterministic and, due to (\ref{eqn:permutation-model2}),
\[
f(R_k | \overline{Y}_{k}^{m}, Y_k, \underline{O}_k^m) = \frac{ \exp \{ R_k ( h_k(\overline{Y}^m_{k}, \underline{O}^m_k;\alpha_k) + \alpha_k Y_k)\}}{ 1+ \exp\{ h_k(\overline{Y}^m_{k}, \underline{O}^m_k;\alpha_k) + \alpha_k Y_k\} }
\]
Thus, (\ref{eqn:dag-f}) is a full data law factorization with respect to a DAG ${\cal G}$ defined using the ordering (\ref{eqn:order}), where for each $k$, the parents of $Y_k$ are $\overline{Y}_k^m$, the parents of $R_k$ are $\overline{Y}_{k}^{m}, Y_k, \underline{O}_k^m$, and the parents of $Y^{obs}_k$ are $R_k,Y_k$.

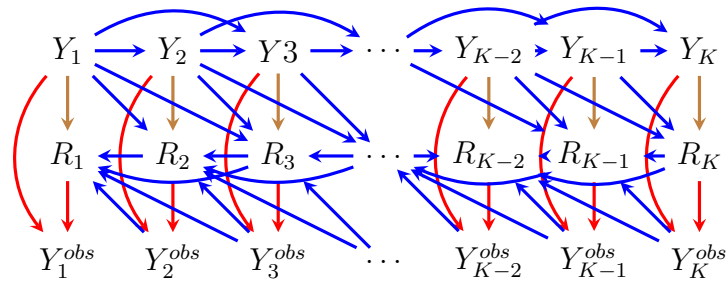
\begin{figure}
	\begin{center}
		\begin{tikzpicture}[>=stealth, node distance=1.4cm]
		\tikzstyle{format} = [draw, very thick, circle, minimum size=5.0mm,
		inner sep=0pt]
		\tikzstyle{unode} = [draw, very thick, circle, minimum size=1.0mm,
		inner sep=0pt]
		\tikzstyle{square} = [draw, very thick, rectangle, minimum size=4mm]
		
		\begin{scope}[xshift=0.0cm]
		\path[->, very thick]
		node[] (x11) {$Y_1$}
		node[right of=x11] (x12) {$Y_2$}
		node[right of=x12] (x13) {$Y_3$}
		node[right of=x13] (dots) {$\ldots$}
		node[right of=dots] (x1k-2) {$Y_{K-2}$}
		node[right of=x1k-2] (x1k-1) {$Y_{K-1}$}
		node[right of=x1k-1] (x1k) {$Y_{K}$}
		
		node[below of=x11] (r1) {$R_1$}
		node[below of=x12] (r2) {$R_2$}
		node[below of=x13] (r3) {$R_3$}
		node[below of=x1k-2] (rk-2) {$R_{K-2}$}
		node[below of=x1k-1] (rk-1) {$R_{K-1}$}
		node[below of=x1k] (rk) {$R_K$}
		
		node[below of=r1] (x1) {$Y_1^{obs}$}
		node[below of=r2] (x2) {$Y_2^{obs}$}
		node[below of=r3] (x3) {$Y_3^{obs}$}
		node[below of=rk-2] (xk-2) {$Y_{K-2}^{obs}$}
		node[below of=rk-1] (xk-1) {$Y_{K-1}^{obs}$}
		node[below of=rk] (xk) {$Y_K^{obs}$}
		
		node[below of=dots] (dots2) {$\ldots$}
		node[below of=dots2] (dots3) {$\ldots$}


		(r1) edge[red] (x1)
		(r2) edge[red] (x2)
		(r3) edge[red] (x3)
		(rk-2) edge[red] (xk-2)
		(rk-1) edge[red] (xk-1)
		(rk) edge[red] (xk)


		(x11) edge[red, bend right=40] (x1)
		(x12) edge[red, bend right=40] (x2)
		(x13) edge[red, bend right=40] (x3)
		(x1k-2) edge[red, bend right=40] (xk-2)
		(x1k-1) edge[red, bend right=40] (xk-1)
		(x1k) edge[red, bend right=40] (xk)

		(x11) edge[brown] (r1)
		(x12) edge[brown] (r2)
		(x13) edge[brown] (r3)
		(x1k-2) edge[brown] (rk-2)
		(x1k-1) edge[brown] (rk-1)
		(x1k) edge[brown] (rk)
		
		(r2) edge[blue] (r1)
		(r3) edge[blue] (r2)
		(dots2) edge[blue] (r3)
		(rk-2) edge[blue] (dots2)
		(rk-1) edge[blue] (rk-2)
		(rk) edge[blue] (rk-1)

		(x2) edge[blue] (r1)
		(x3) edge[blue] (r2)		
		(dots3) edge[blue] (r3)
		(xk-2) edge[blue] (dots2)
		(xk-1) edge[blue] (rk-2)
		(xk) edge[blue] (rk-1)
		
		(x11) edge[blue] (r2)
		(x12) edge[blue] (r3)
		(x13) edge[blue] (dots2)
		(x1k-2) edge[blue] (rk-1)
		(x1k-1) edge[blue] (rk)
		
		(x1k-1) edge[blue] (x1k)
		(x1k-2) edge[blue] (x1k-1)
		(dots) edge[blue] (x1k-2)
		(x13) edge[blue] (dots) 
		(x12) edge[blue] (x13)
		(x11) edge[blue] (x12)
		
		
		(x11) edge[blue, bend left] (x13)
		(x12) edge[blue, bend left] (dots)
		(dots) edge[blue, bend left] (x1k-1)
		(x1k-2) edge[blue, bend left] (x1k)
		
		(x3) edge[blue] (r1)
		(dots3) edge[blue] (r2)
		(xk-1) edge[blue] (dots2)
		(xk) edge[blue] (rk-2)
		
		(r3) edge[blue, bend left=20] (r1)
		(dots2) edge[blue, bend left=20] (r2)
		(rk-1) edge[blue, bend left=20] (dots2)
		(rk) edge[blue, bend left=20] (rk-2)
		
		(x11) edge[blue] (r3)
		(x12) edge[blue] (dots2)
		(dots) edge[blue] (rk-1)
		(x1k-2) edge[blue] (rk)

		;
		\end{scope}

		\end{tikzpicture}
	\end{center}
	\caption{
	 Second-order Markov submodel. Red edges indicate a deterministic relationship between the parents of the variable (in this case the deterministic relationship between $Y_i^{obs}$ and parents $R_i, Y_i)$. Brown edges indicate edges introduced through non-zero $\alpha$ in exponential tilting for sensitivity analysis. Blue edges indicate a non-deterministic probabilistic dependence of a variable on its parents (e.g. a standard DAG edge).
	}
	\label{fig1}
\end{figure}

As an example, Figure~\ref{fig1} is a DAG that represents our model when $m=2$, with arrows into each $Y^{obs}_k$ are shown in red to emphasize that the relationship between $Y^{obs}_k$ and its parent variables $R_k,Y_k$ is deterministic.
The set of parents of all other variables in the model correspond to restrictions (\ref{eqn:permutation-model0a}), (\ref{eqn:permutation-model0b}), and (\ref{eqn:permutation-model2}).  We display the arrow from $Y_k$ to $R_k$ for each $k$ in brown, to emphasize the fact that dependence of $R_k$ on $Y_k$ in the conditional distribution $f(R_k \mid \overline{Y}^m_k, Y_k, \underline{O}^m_k)$ is parameterized by the sensitivity parameter $\alpha_k$, and this dependence disappears if $\alpha_k = 0$.  

As discussed in \cite{scharfstein2021global}, the distribution of $\overline{Y}$ with $m=K-1$ (no Markovian restrictions) is identified.  This means, for example, that we can express $P(Y_k=y_k|Y_{K-1}=y_{K-1},Y_{K-2}=y_{K-2}, Y_{K-3}=y_{K-3})$ as a (complicated) functional, $g(y_K,y_{K-1},y_{K-2},y_{K-3})$, of the distribution of the observed data. To see why the Markovian restrictions on $f(\overline{Y})$ impose restrictions on the observed data law, consider the model Figure~\ref{fig1}.  Here, $Y_K \ci Y_{K-3} \mid Y_{K-2},Y_{K-1}$. This implies that $g(y_K,y_{K-1},y_{K-2},y_{K-3})$ cannot depend on $y_{K-3}$ which has testable implications.





\begin{aside}
\noindent {\bf An Equivalent Model Obeying A Total Order Induced By Temporal Precedence:} A critique of DAGs representing missing data models is that, unlike DAGs representing causal models, they may contain arrow orientations that do not obey the temporal order on variables.  This can complicate the interpretability of such models.
Our proposed model of the full data does not suffer from this issue, in the following sense.  While it is represented by a DAG where some arrows must necessarily not be consistent with the temporal order on indices of variables, our model can always be represented by a mixed graph (i.e., directed and undirected edges) representing a DAG on the full data where all arrows are consistent with a temporal order.

A well known result states that two DAGs imply the same Markov model if they agree on the presence or absence of edges (ignoring orientations), and on all \emph{unshielded colliders}\citep{verma90equiv}.  An unshielded collider is a structure of the form $A \to B \gets C$, where $A$ and $C$ are not adjacent.  Given this result, we see that a DAG ${\cal G}^*$ equivalent to the DAG in Fig.~\ref{fig1} is one where all arrows among $Y_1, \ldots, Y_K$ are reversed.  This is because this change does not change the presence of edges, or unshielded colliders in the DAG.

It is possible to replace edges of the form $Y_k \to R_{k+1}$ and $Y_k \to R_{k+2}$ by bidirected edges $Y_k \leftrightarrow R_{k+1}$ and $Y_k \leftrightarrow R_{k+2}$ signifying the presence of unobserved common parents:
$Y_k \gets U_{k+1} \to R_{k+1}$ and $Y_k \gets U_{k+2} \to R_{k+2}$, where $U_{k+1}$ and $U_{k+2}$ are unmeasured or hidden variables.  This hidden variable DAG $\tilde{\cal G}$ implies the same model as ${\cal G}^*$ since, as before, the set of vertex adjacencies and unshielded colliders have not changed.  Furthermore, DAG $\tilde{\cal G}$ has a total order on variables that is consistent with a temporal order on indices where larger indices occur earlier than smaller ones.  

Importantly, a similar construction is possible for a DAG representing any $m$-order Markov model we consider.
\end{aside}


\section{Identification}


In this section, we show that the full data distribution is identified in our model, provided, for $k=1,\ldots,K$, $P(\overline{O}_{k}^m = \overline{o}_{k}^m, \underline{O}_{k-1}^{m+1}=\underline{o}_{k-1}^{m+1} )>0$ for all $\overline{o}_{k}^m$ and $\underline{o}_{k-1}^{m+1}$. 
First, we use the fact that our full data distribution obeys the global Markov property with respect to a DAG to prove (see Appendix) the following two results.
\begin{restatable}{lem}{firstind}\label{lem:ind1}
For $1 < k \leq K-m -1$, $\overline{R}^m_k$  is independent of $O_{k+m+1}$ given $\overline{Y}_{k}^{m},\underline{O}^{m+1}_{k-1}$.
\end{restatable}
\begin{restatable}{lem}{secondind}\label{lem:ind2}
For $1 \leq k \leq K-m -1$, $R_k$ is independent of $O_{k+m+1}$ given $\overline{Y}_{k+1}^{m+1},\underline{O}^{m}_{k}$.
\end{restatable}
Then, we prove by induction (see Appendix) the following identification result:
\begin{restatable}{lem}{idinduction}\label{lem:id1}
For $k=1,\ldots, K$, $f(\overline{Y}_{k+1}^{m+1},\underline{O}^{m+1}_{k})$ is identified.  Specifically,
\begin{align}
f(\overline{Y}_{k+1}^{m+1},\underline{O}^{m+1}_{k}) &=  f(O_{k+m+1} | R_k=1, \overline{Y}_{k+1}^{m+1}, \underline{O}^{m}_k) ^{I(k \leq K - m - 1)} f(Y_k|R_k=1,\overline{Y}_{k}^{m}, \underline{O}^{m}_k)
\nonumber \\
&\quad \left\{ P(R_k=1|\overline{Y}_{k}^{m},\underline{O}^{m}_{k}) + P(R_k=0|\overline{Y}_{k}^{m},\underline{O}^{m}_{k}) \frac{ \exp\{ \alpha_k Y_k \}}{c_k(\overline{Y}_{k}^{m},\underline{O}^{m}_{k};\alpha_k) } \right\} f(\overline{Y}_{k}^{m},\underline{O}^{m}_{k} ),
\label{eqn:margin-id1}
\end{align}
where the right hand side of (\ref{eqn:margin-id1}) is a function of $f(\overline{Y}_{k}^{m},\underline{O}^{m+2}_{k-1} )$, which is inductively identified because 
\begin{eqnarray} \label{eqn:margin-id2}
f(\overline{Y}_{k}^{m},\underline{O}^{m+2}_{k-1} ) &= & \left\{ f(O_{k+m+1} | \overline{Y}^m_{k},\underline{O}^{m+1}_{k-1}, \overline{R}_k^m=1 ) \right\}^{I(k \leq K-m-1)} \times \nonumber\\
& & \left\{ f(\overline{Y}_{k}^{m+1},\underline{O}^{m+1}_{k-1} ) \right\}^{I(k \leq m+1)} \left\{ \sum_{Y_{k-m-1}} f(\overline{Y}_{k}^{m+1},\underline{O}^{m+1}_{k-1} ) \right\}^{I(k > m+1)} \label{eqn:marginal-id2}
\end{eqnarray}

\end{restatable}

Lemma \ref{lem:id1} immediately implies identification of $f(\overline{Y})$ under our model.

\begin{thm} \label{thm:target_law}
$f(\overline{Y})$ is identified as $\prod_{k=1}^{K} f(Y_k | \overline{Y}^m_k) = \prod_{k=1}^{K} \left\{ \frac{ f(\overline{Y}^{m+1}_{k+1}) }{ \sum_{Y_k} f(\overline{Y}^{m+1}_{k+1}) } \right\}$,
where\\ $f(\overline{Y}_{k+1}^{m+1}) =  \sum_{\underline{O}_{k}^{m+1}} f(\overline{Y}_{k+1}^{m+1}, \underline{O}_{k}^{m+1})$.
\end{thm} 


In the Appendix, we illustrate the above lemmas and theorem for $K=7$ and $m=2$.





\section{Inference}\label{sec:inference}

The identification result of Lemma \ref{lem:id1} and Theorem \ref{thm:target_law} shows us that there is a $\alpha_k$-dependent forward time inductive mapping from the distribution of the observed data to
$f(Y_k| \overline{Y}_k^m)$ and $P(R_{k}=1| \overline{Y}_{k}^{m}, \underline{O}_{k}^{m} )$ for $k=1,\ldots,K$.  At step $k$, the mapping depends on $f(\overline{O}_k^m,\underline{O}_{k-1}^{m+2})= f(\underline{O}_{k-m-1}^{2m+2})= f(\overline{O}_{k+m+2}^{2m+2})$.
It is these distributions that we will estimate (see below). 
In the Appendix, we show that there is a $\alpha_k$-dependent reverse time inductive mapping from  $f(Y_k| \overline{Y}_k^m)$ and $P(R_{k}=1| \overline{Y}_{k}^{m}, \underline{O}_{k}^{m} )$ for $k=1,\ldots,K$ to a model-respecting distribution of observed data.  As discussed above, our model places complicated restrictions on the distribution of the observed data that are hard to model directly. In fact, our estimator of $f(\overline{O}_k^m,\underline{O}_{k-1}^{m+2})$ may not cohere with the model restrictions.


We estimate the distributions $f(\overline{O}_{k+m+2}^{2m+2})$, $k=1,\ldots,K$, using Laplacian smoothing, where the smoothing parameter is selected by cross-validation.  Specifically, let $\widehat{P}(\overline{O}=\overline{o})$ be the empirical probability of the event $(\overline{O}=\overline{o})$, where $\overline{o}$ is a potential realization of $\overline{O}$. Let $\widetilde{P}(\overline{O}=\overline{o};\delta) = \frac{ \widehat{P}(\overline{O}=\overline{o}) + \lambda}{1+3^K \lambda}$, where $\lambda$ is a smoothing parameter.  Storing and operating on $\widetilde{P}(\overline{O}=\overline{o};\lambda)$ defeats the purpose of the Markov restrictions, as the space complexity of this distribution scales exponentially in $K$. However, this can be avoided by noting that the marginalization of  $\widetilde{P}(\overline{O}=\overline{o};\lambda)$ implies that 
$$\widetilde{P}(\overline{O}_{k'}^{j'} = \overline{o}_{k'}^{j'};\lambda) =
\frac{\widehat{P}(\overline{O}_{k'}^{j'} = \overline{o}_{k'}^{j'}) + 3^{h(k',j')}  \lambda }{1+3^K \lambda },$$
where $h(k',j') = \max\{k'-j'-1,0\}+ K - k' + 1$. We need to compute $\widetilde{P}(\overline{O}_{k'}^{j'} = \overline{o}_{k'}^{j'};\lambda)$ for $k'=k+m+2$ and $j'=2m+2$, $k=1,\ldots,K$.  Thus, we only require storage of the empirical distribution of $\overline{O}_{k+m+2}^{2m+2}$, which has size complexity $3^{k+m+1-\max\{k-m-1,0\}}$. This size complexity is exponential in $m$ rather than $K$.

We use cross-validation to estimate $\lambda$. Specifically, we split the dataset randomly into $L$ parts. Let ${\cal S}_l$ and ${\cal S}_{-l}$ be the set of individuals in and not in the $l$th split, respectively. Let $\widehat{P}_l(\cdot)$ and $\widehat{P}_{-l}(\cdot)$ be the empirical distributions of based in ${\cal S}_l$ and ${\cal S}_{-l}$, respectively. Similarly, let $\widetilde{P}_l(\cdot;\lambda)$ and $\widehat{P}_{-l}(\cdot;\lambda)$ be the associated smoothed distributions with smoothing parameter $\lambda$. We estimate $\lambda$ by $\widetilde{\lambda}$ to be the minimizer of the following cross-validated loss function:
\[
{\cal L}(\lambda) = \sum_{l=1}^L \sum_{k=1}^K \sum_{\overline{o}_{k+m+2}^{2m+2}} \{ \widehat{P}_l(\overline{O}_{k+m+2}^{2m+2} = \overline{o}_{k+m+2}^{2m+2}) - \widetilde{P}_{-l}(\overline{O}_{k+m+2}^{2m+2} = \overline{o}_{k+m+2}^{2m+2};\lambda) \}^2.
\]
We then estimate $P(\overline{O}_{k+m+2}^{2m+2} = \overline{o}_{k+m+2}^{2m+2})$ by 
$\widetilde{P}(\overline{O}_{k+m+2}^{2m+2} = \overline{o}_{k+m+2}^{2m+2};\widetilde{\lambda})$. Let $\widetilde{f}(\overline{O}_{k+m+2}^{2m+2};\widetilde{\lambda})$ represent our estimator of $f(\overline{O}_{k+m+2}^{2m+2})$. The work of \cite{grund1993kernel} suggest that our estimators of $f(\overline{O}_{k+m+2}^{2m+2})$ will converge at $\sqrt{n}$-rates and be jointly asymptotically normal. 

In data analysis and simulation study, we consider the target parameter of interest to be the $\psi = E[\sum_{k=1}^{K} Y_k] = \sum_{k=1}^K E[Y_k]$, where $K=24$. For given $\alpha_k$'s, $\psi$ is identified as a smooth, but complicated functional of ${\cal F} = \{ f(\overline{O}_{k+m+2}^{2m+2}): k=1,\ldots,K \}$. Suppressing dependence on $\alpha_k$'s, let $\psi({\cal F})$ denote this functional.  It is natural to consider a plug-in estimator of $\psi$, i.e., $\widetilde{\psi}_{\mbox{p-i}} = \psi(\widetilde{{\cal F}})$, where $\widetilde{{\cal F}} = \{ \widetilde{f}(\overline{O}_{k+m+2}^{2m+2};\widetilde{\lambda}): k=1,\ldots,K \})$.  In simulations, we found that this estimator has excessive bias for the typical sample sizes seen in trials like CTN-0044. To address this issue, we use a bias-corrected or one-step estimator, which is equal to plug-in estimator plus the average of individual-specific influence functions \citep{bickel1993efficient}. Given the complexity of recursive nature of identification algorithm, it is hard to compute a closed-form expression for the influence functions. Thus, we use a numerical approximation as discussed by \cite{frangakis2015deductive, carone2019toward, jordan2022data}.  Specifically, we approximate the influence function for person $i$ by 
\[
\widetilde{\rho}(\overline{O}_i) = \frac{ \psi(\widetilde{{\cal F}_{-i}}+\epsilon \delta(\overline{O}_i)) - \psi(\widetilde{{\cal F}_{-i}}) }{\epsilon},
\]
where $\widetilde{{\cal F}_{-i}}$ is the smoothed estimate of ${\cal F}$ based on a dataset where the $i$th individual is removed, $\delta(\cdot)$ is the Dirac-delta function, and $\epsilon$ is small positive number. Our bias-corrected estimator is:
\[
\widetilde{\psi}_{\mbox{b-c}} = \widetilde{\psi}_{\mbox{p-i}} + \frac{1}{n} \sum_{i=1}^{n} \widetilde{\rho}(\overline{O}_i)
\]
Since $\psi(\cdot)$ is a smooth function, one can use Taylor series expansion arguments to show that $\sqrt{n} \{ \widetilde{\psi}_{\mbox{b-c}}  - \psi({\cal F}) \}$ converges to a normal distribution.

We construct confidence intervals using parametric bootstrap.  Specifically, for each specified set of $\alpha_k$'s, we repeat the following procedure $B$ times.
\begin{itemize}
\item Use $\widetilde{{\cal F}}$ to generate a dataset of size $n$ (see simulation procedure in Appendix 8.2).
\item On the generated dataset, estimate ${\cal F}$ using the smoothing procedure and estimate $\psi$ using the bias-corrected estimation procedure.
\end{itemize}
We then compute percentile bootstrap confidence interval based on the $B$ estimates of $\psi$.

\section{Analysis of CTN-0044}

In CTN-0044, individuals were randomized to 12 weeks of either treatment-as-usual (TAU, $n=252$) or treatment-as-usual plus a computerized psychosocial intervention with contingent incentives  (TAU+, $n=255$).  At each half-week, an individual was considered to be abstinent
if the urine screen was negative and the self-report indicated no drug use/any alcohol use, and not abstinent otherwise. If self-report was missing, but urine screen was positive, the half-week was scored as not abstinent. Abstinence during a given half-week was considered missing if (a) self-report indicated no use but the urine screen was missing, (b) the urine screen was negative but self-report was missing, or (c) both urine and self-report were missing. Table \ref{tab0} summarizes missingness patterns by treatment group.  The table shows high rates of non-monotone missing data in both arms and lower rates of missing data for the TAU+ arm.
 
For each treatment group, we are interested in drawing inference about the mean number of abstinent (i.e., negative) urine samples, i.e., $E[ \sum_{k=1}^{24} Y_k] = \sum_{k=1}^{24} E[Y_k]$, where $Y_k=1$ denotes an abstinent measurement at time $k$.  
Based on our simulation results (next section) and the given sample size, we only considered a first-order Markovian model (i.e., $m=1$).


\begin{table}
\caption{CTN-0044: Missingness Patterns}
\begin{center}
\begin{tabular}{lcc}
Missingness Pattern & TAU ($n=252$) & TAU+ ($n=255$) \\ \hline
Complete & 42 (16.7\%)  & 66 (25.9\%)  \\
Monotone \\
\; \; $1,2,\ldots,23$ Missing & 10 (4.0\%) &  13 (5.1\%)  \\
\; \; All Missing &  7 (2.8\%)  & 4 (1.6\%)  \\
Non-monotone \\ 
 \; \; $1$ Missing & 23 (9.1\%) & 27 (10.6\%) \\    
 \; \;  $2$ Missing & 35 (13.9\%) & 37 (14.5\%) \\  
 \; \;  $3$ Missing & 17 (6.8\%) & 15 (5.9\%) \\  
 \; \;  $4$ Missing & 13 (5.2\%) & 11 (4.3\%) \\  
 \; \;  $5$ Missing & 15 (6.0\%) & 13 (5.1\%) \\ 
 \; \;  $6$ Missing & 15 (6.0\%) & 8 (3.1\%) \\ 
 \; \;  $7$ Missing & 7 (2.8\%) & 9 (3.5\%) \\ 
 \; \;  $8$ Missing & 12 (4.8\%) & 11 (4.3\%) \\ 
 \; \;  $9$ Missing & 7 (2.8\%) & 6 (2.4\%) \\ 
 \; \;  $10$ Missing & 7 (2.8\%) & 6 (2.4\%) \\ 
 \; \;  $11$ Missing & 6 (2.4\%) & 3 (1.2\%) \\ 
 \; \;  $12$ Missing & 7 (2.8\%) & 1 (0.4\%) \\ 
\; \;  $13$ Missing & 3 (1.2\%) & 9 (3.5\%) \\ 
 \; \;  $14$ Missing & 7 (2.8\%) & 4 (1.6\%) \\ 
 \; \;  $15$ Missing & 4 (1.6\%) & 3 (1.2\%) \\ 
 \; \;  $16$ Missing & 3 (1.2\%) & 2 (0.8\%) \\ 
\; \;  $17$ Missing & 6 (2.4\%) & 1 (0.4\%) \\ 
\; \;  $18$ Missing & 1 (0.4\%) & 5 (2.0\%) \\ 
\; \;  $19$ Missing & 1 (0.4\%) & 0 (0.0\%) \\ 
\; \;  $20$ Missing & 2 (0.4\%) & 0 (0.0\%) \\ 
\; \;  $21$ Missing & 0 (0.4\%) & 0 (0.0\%) \\ 
\; \;  $22$ Missing & 1 (0.4\%) & 0 (0.0\%) \\ 
\; \;  $23$ Missing & 1 (0.4\%) & 1 (0.4\%) \\ 
\hline
\end{tabular}
\end{center}
   \label{tab0}
\end{table}


For each treatment group, we estimated the average number of abstinent half-weeks under the following assumptions: missing completely at random (MCAR), missing equals abstinent, missing equals not abstinent, and under  $\alpha_k=0$.  Table \ref{tab1} displays the treatment-specific estimates and difference in estimates, along with 95\% bootstrap confidence intervals (parametric bootstrap for benchmark analysis - 500 samples; non-parametric bootstrap for other analyses - 1000 samples). While all treatment effect estimates favored TAU+, only the confidence interval for missing equals abstinent included zero. Relative to MCAR, the treatment-specific estimates of the mean number of abstinent samples was higher under the benchmark assumption, indicating that the missing samples were more likely (under the benchmark assumption) to be abstinent than those observed. This seems counter-intuitive as prevailing wisdom suggests that missing samples would correspond to non-abstinent drug behavior. However, the outcome involves the results of a urine drug test {\em and} self-reported substance use. The outcome is considered missing if the urine test is negative, but the participant fails to self-report. In contrast, the participant is considered non-abstinent if their urine test is positive, regardless of self-report. The study protocol provided specific financial compensation for providing half-week urine samples and for detailed in-person surveys at 4, 8 and 12 weeks.   These surveys were designed to collect detailed information on risk behaviors, social adjustment, psychiatric symptoms, coping strategies and self-reported substance use over the past 4 weeks. Compensation for these surveys was not contingent on completing the self-report form. Thus, a participant who would have negative urine at a half-week has a clear incentive to provide a urine sample. While a participant who would have positive urine at a half-week has a financial incentive to provide a urine sample, this incentive may be offset by their substance use behavior.  Neither participant has an significant incentive to complete the self-report form of the subsequent survey.  Thus, there are competing factors at play that influence missingness of the half-week abstinent measure.  These competing factors make the comparison of the MCAR and benchmark results more subtle than prevailing wisdom may suggest.







\begin{table}
\caption{Inference under missing completely at random (MCAR), missing equals abstinent and missing equals not abstinent as well as the benchmark assumption ($\alpha_k=0$, $m=1$).}
\begin{center}
\begin{tabular}{llll}
Assumption  &     TAU &                   TAU+         &            Difference \\ \hline
MCAR & 11.72 (10.54, 13.15) & 13.99 (12.83, 15.07) & 2.26 (0.35, 3.90) \\
Missing=Abstinent & 14.82 (13.91, 15.83) & 16.04 (15.05, 17.02) & 1.23 (-0.26, 2.59) \\
Missing=Not Abstinent & 8.80 (7.79, 9.83) & 11.10 (9.92, 12.23) & 2.30 (0.72, 3.75) \\
Benchmark ($\alpha_k=0$) & 12.31 (11.36, 13.14) & 14.29 (13.39, 15.02) & 1.93 (1.04, 2.75) \\
\hline
\end{tabular}
\end{center}
\label{tab1}
\end{table}

\begin{figure}[!tbp]
  \begin{subfigure}[b]{0.5\textwidth}
    \includegraphics[width=\textwidth]{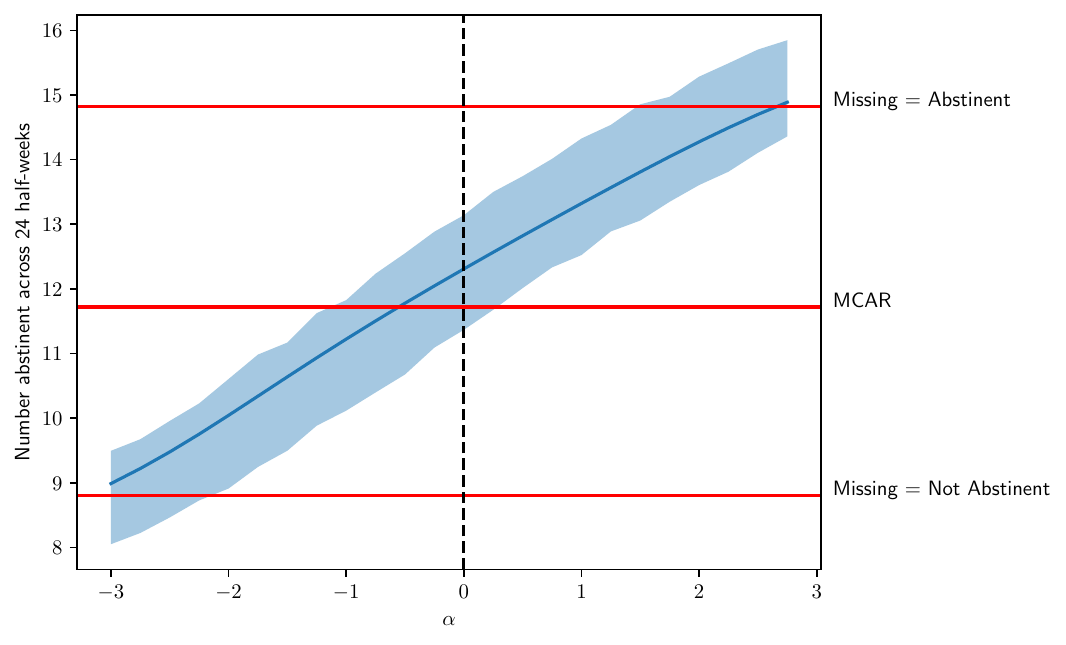}
    \caption{TAU}
  \end{subfigure}
  \hfill
  \begin{subfigure}[b]{0.5\textwidth}
    \includegraphics[width=\textwidth]{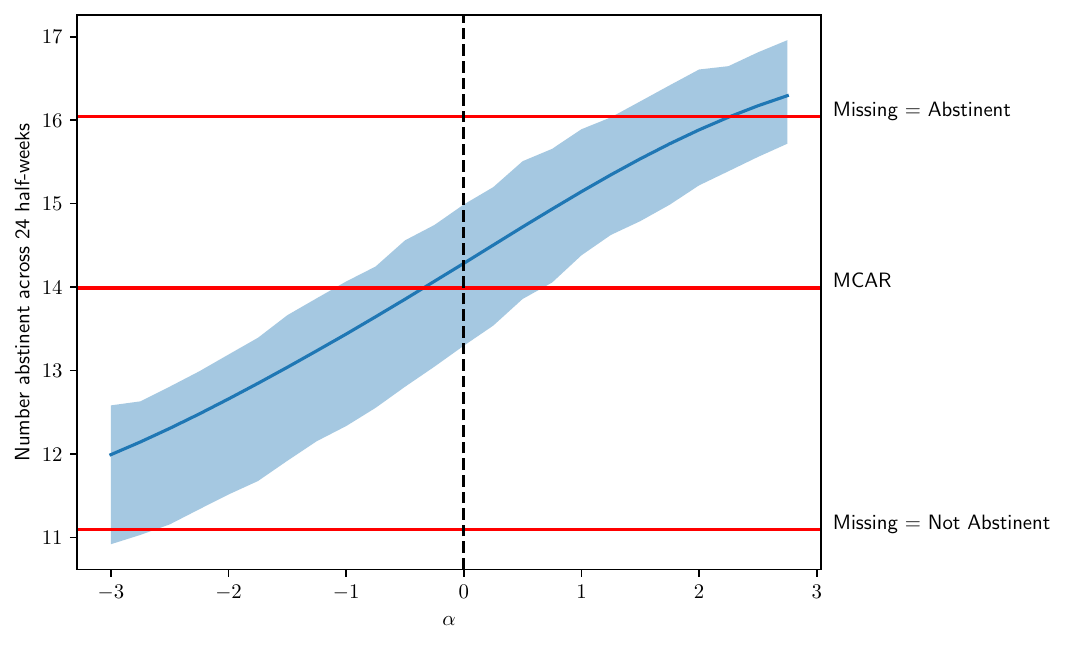}
    \caption{TAU+}
  \end{subfigure}
  \caption{Treatment-specific estimates of mean number of abstinent half-weeks (along with 95\% confidence intervals) as a function of $\alpha$. This figure appears in color in the electronic version of this article, and any mention of color refers to that version.}
    \label{fig:f1}
\end{figure}

\begin{figure}[!tbp]
    \includegraphics[width=0.9\textwidth]{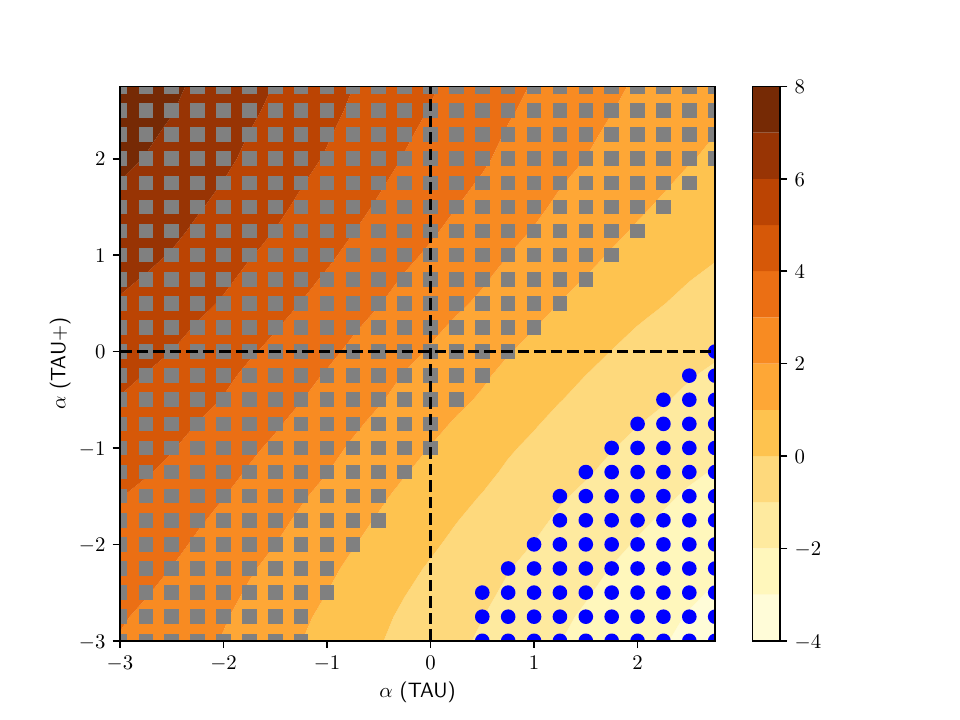}
    \caption{A contour plot of estimated treatment differences, as a function of treatment-specific sensitivity analysis parameters. Combinations of treatment-specific sensitivity parameters with a blue circle or gray square indicate that the associate 95\% confidence interval excludes 0.  Treatment differences marked by a gray square dot correspond to positive treatment effects where TAU+ is favored, and differences marked by a blue dot correspond to negative treatment effects where TAU is favored. This figure appears in color in the electronic version of this article, and any mention of color refers to that version.}
    \label{fig:f2}
\end{figure}

In our sensitivity analysis, we assumed $\alpha_k=\alpha$ over all $k$.  For each treatment group and each $\alpha$ (ranging from -3 to 3), we estimated the mean number of abstinent half weeks.  The results 
(along with 95\% percentile parametric bootstrap confidence intervals - 500 samples)
are presented in Figure \ref{fig:f1}.  
Figure \ref{fig:f2} is a contour plot which shows that treatment effect estimates for various combinations of treatment-specific sensitivity analysis parameters.  The figure also displays regions where there evidence of an effect in favor of TAU+ (gray squares) and in favor of TAU (blue dots).  While it is impossible to determine the true value of the treatment-specific sensitivity analysis parameters, researchers can use their scientific judgment to narrow the range of these parameters. One might expect that the factors that drive missingness are common across treatment arms.  This would imply that the sign of the treatment-specific sensitivity parameters will be the same and the magnitudes would be similar. This suggests a focus on a strip around the 45 degree line that intersects the lower left or the upper right quadrants of Figure \ref{fig:f2}.  If the strip is of width 1.0, then all point estimates of the treatment effect favor TAU+, but there are values of the sensitivity analysis parameters that yield confidence intervals including 0.0 (e.g., $\alpha(\mbox{TAU})=1.0$ and  $\alpha(\mbox{TAU+})=0.0$. This does suggest sensitivity of inferences to deviations from the benchmark assumption. However, with these beliefs, there is no evidence supportive of TAU over TAU+.

\section{Simulation Study}

For each treatment group, we used the CTN-0044 to develop a realistic simulation study. To start, we specified $m$ and used the observed data to estimate $f(\overline{O}_{k+m+2}^{2m+2})$ by $\widetilde{f}(\overline{O}_{k+m+2}^{2m+2};0.2)$, $k=1,\ldots, K$.  Using $\alpha_k=0$, we then used our algorithm to estimate $P(Y_k=1| \overline{Y}_{k}^{m})$ and $P(R_{k}=1| \overline{Y}_{k}^{m}, \underline{O}_{k}^{m} )$, $k=1,\ldots, K$.  We treat these distributions as the truth.  We can compute the true mean number of abstinent half-weeks using $P(Y_k=1| \overline{Y}_{k}^{m})$.  We can use the algorithm in Section 8.2 of the Appendix to generate datasets for any specified $n$ and choice of $\alpha_k$'s.  In our simulation study, we considered $m=1,2$, $n=100, 250, 500, 1000$ and $\alpha_k=\alpha=-3,-2,-1,0,1,2,3$.  We evaluated bias of our bias-corrected estimator and coverage of 95\% confidence intervals.

Table \ref{tab:tab3} summarizes our simulation results. For each treatment group, $m$ and $\alpha$, bias decreases with $n$. For $m=1$, there is under-coverage at $n=100$, but for all
$n \geq 250 $, the coverage is at or close to the 95\% level. By contrast, for $m=2$, sizes of $n=1000$ are needed to achieve 95\% coverage. This is expected, as the $m=2$ model has roughly 4 times the number of parameters, and so requiring 4 times the amount of data is not surprising. 
\begin{table}
\caption{Results of simulation study}
\label{tab:tab3}
\begin{tabular}{rrrr|rr|rr|rr}
\toprule
  & & \multicolumn{2}{c}{TAU, $m=1$} & \multicolumn{2}{c}{TAU+, $m=1$} & \multicolumn{2}{c}{TAU, $m=2$} & \multicolumn{2}{c}{TAU+, $m=2$} \\
\toprule
  Truth: & & \multicolumn{2}{c}{$12.24$} & \multicolumn{2}{c}{$14.08$} & \multicolumn{2}{c}{$12.20$} & \multicolumn{2}{c}{$14.00$} \\
\toprule
$\alpha$ & $n$ & Mean & Coverage & Mean & Coverage & Mean & Coverage & Mean & Coverage\\
\midrule
-2 & 100 & 11.62 & 0.85  & 13.73 & 0.93 &  10.98 & 0.20 & 13.18 & 0.46 \\  
-2 & 250 & 11.95 & 0.94  & 13.91 & 0.94 &  11.55 & 0.59 & 13.67 & 0.85 \\
-2 & 500 & 12.11 & 0.95  & 13.99 & 0.96 &  11.86 & 0.82 & 13.88 & 0.95 \\
-2 & 1000 & 12.17 & 0.94 & 14.03 & 0.94 &  12.05 & 0.94 & 13.97 & 0.95 \\
\hline
-1 & 100 & 11.93 & 0.96  & 13.87 & 0.94 &  11.49 & 0.68 & 13.46 & 0.71 \\   
-1 & 250 & 12.09 & 0.93  & 13.97 & 0.94 &  11.84 & 0.84 & 13.83 & 0.94 \\    
-1 & 500 & 12.18 & 0.96  & 14.02 & 0.95 &  12.05 & 0.92 & 13.95 & 0.96 \\ 
-1 & 1000 & 12.21 & 0.95 & 14.04 & 0.96 &  12.16 & 0.96 & 13.99 & 0.94 \\ 
\hline
0 & 100 & 12.17 & 0.98   & 13.99 & 0.96 &  12.11 & 0.96 & 13.88 & 0.92 \\   
0 & 250 & 12.21 & 0.96   & 14.02 & 0.95 &  12.14 & 0.96 & 13.98 & 0.96 \\   
0 & 500 & 12.24 & 0.95   & 14.05 & 0.95 &  12.17 & 0.96 & 14.00 & 0.95 \\   
0 & 1000 & 12.24 & 0.95  & 14.06 & 0.95 &  12.19 & 0.95 & 14.00 & 0.95 \\   
\hline
1 & 100 & 12.44 & 0.97   & 14.15 & 0.96 &  12.73 & 0.78 & 14.32 & 0.93 \\  
1 & 250 & 12.32 & 0.95   & 14.09 & 0.95 &  12.42 & 0.92 & 14.22 & 0.90 \\  
1 & 500 & 12.29 & 0.95   & 14.07 & 0.94 &  12.31 & 0.95 & 14.11 & 0.92 \\  
1 & 1000 & 12.25 & 0.95  & 14.07 & 0.93 &  12.22 & 0.95 & 14.00 & 0.97 \\   
\hline
2 & 100 & 12.72 & 0.92   & 14.33 & 0.94 &  13.30 & 0.30 & 14.69 & 0.70 \\  
2 & 250 & 12.43 & 0.95   & 14.17 & 0.95 &  12.73 & 0.66 & 14.43 & 0.76 \\  
2 & 500 & 12.33 & 0.96   & 14.12 & 0.96 &  12.50 & 0.87 & 14.22 & 0.86 \\  
2 & 1000 & 12.28 & 0.95  & 14.08 & 0.96 &  12.31 & 0.95 & 14.06 & 0.94 \\   
\bottomrule
\end{tabular}
\end{table}



\section{Discussion}

In this paper, we re-visited the sensitivity analysis model introduced by \cite{scharfstein2021global} for the analysis of randomized trials with longitudinal binary ouctcomes subject to nonmonotone missingness.  We imposed an $m$th-order Markovian restriction that makes it computationally feasible to handle studies designed to collect a large number (e.g., greater than 15) of outcomes on each individual. 

The Markov restriction on the full data distribution was imposed in a coherent way by using the factorization with respect to a directed acyclic graph (DAG).  This factorization allowed a number of additional conditional independence restrictions on the full data law to be easily derived using the d-separation criterion \citep{pearl88probabilistic}.  It would have been difficult to derive these conditional independence restrictions without use of graphical models. Importantly, these restrictions allowed us to obtain model identification in a way that led  directly to a tractable inferential strategy. The estimation technique employed Laplacian smoothing, which was implemented to ensure compatibility with the Markov restrictions. 

Techniques based on graphical models used in this paper are very general. An important area of future work is to consider what other tractable sensitivity analysis procedures can be developed using our approach.


\section*{Acknowledgements}

This research was supported by R01DA046534. 

\section*{Data Availability Statement}

The National Drug Abuse Treatment Clinical Trials Network (CTN) protocol \#0044 trial data that were analyzed in this paper are publicly available at the National Institute on Drug Abuse Data Share Website at \url{https://datashare.nida.nih.gov/study/nida-ctn-0044} \citep{ctn0044}.

\bibliographystyle{biom}
\bibliography{refs1}

\newpage 



\newpage 

\section*{Appendix}

\firstind*
\begin{proof}
We consider three cases: (a) $m < k < K - m - 1$, (b) $k \leq m$, and (c) $k = K - m - 1$.

We mark deterministic functions in {\color{red} red text}. For case (a), note that the full data law (\ref{eqn:dag-f}) can be written as
\begin{align}
& \underbrace{ \left(\prod_{j=1}^{k-m-1} f(Y_j | \overline{Y}_j^m) \right) }_{ f(\overline{Y}_{k-m})  } \underbrace{ \left( \prod_{j=k-m}^{k-1} f(Y_j | \overline{Y}_j^m) \right) }_{f(\overline{Y}_{k}^m | \overline{Y}_{k-m})   } \underbrace{ \left(\prod_{j=k}^{k+m} f(Y_j | \overline{Y}_j^m) \right) }_{ f(\underline{Y}_{k-1}^{m+1} | \overline{Y}_{k}^m)  } \nonumber \\
& \underbrace{ f(Y_{k+m+1} | \overline{Y}_{k+m+1}^m) }_{ f(Y_{k+m+1}| \underline{Y}_{k-1}^{m+1}) } \underbrace{ \left(\prod_{j=k+m+2}^{K} f(Y_j | \overline{Y}_j^m) \right)}_{f(\underline{Y}_{k+m+1} |\underline{Y}_{k-1}^{m+2}  ) } \nonumber \\
& \underbrace{ \left(\prod_{j=1}^{k-m-1} f(R_j | \overline{Y}_{j+1}^{m+1}, \underline{O}_j^m) \right)}_{ f(\overline{R}_{k-m} | \overline{Y}_{k-m}, \overline{O}_k^m ) } 
\underbrace{ \left( \prod_{j=k-m}^{k-1} f(R_j | \overline{Y}_{j+1}^{m+1}, \underline{O}_j^m) \right)}_{ f(\overline{R}^m_{k} | \overline{Y}_{k}, \underline{O}_{k-1}^{m+1} )} \underbrace{ \left(\prod_{j=k}^{k+m} f(R_j | \overline{Y}_{j+1}^{m+1}, \underline{O}_j^m) \right)}_{f(\underline{R}_{k-1}^{m+1} | \overline{Y}^{2m+1}_{k+m+1},\underline{O}_{k+m}) } \nonumber \\
& \underbrace{ f(R_{k+m+1} | \overline{Y}_{k+m+2}^{m+1}, \underline{O}_{k+m+1}^m)}_{ f(R_{k+m+1}| \underline{Y}_{k-1}^{m+2},\underline{O}_{k+m+1}) } \underbrace{ \left(\prod_{j=k+m+2}^{K} f(R_j | \overline{Y}_{j+1}^{m+1}, \underline{O}_j^m)\right) }_{ f(\underline{R}_{k+m+1}| \underline{Y}_{k-1})} \nonumber  \\
& \underbrace{ \left(\prod_{j=1}^{k-m-1} {\color{red} f(Y^{obs}_j | R_j, Y_j) } \right)}_{{\color{red} f(\overline{Y}^{obs}_{k-m} | \overline{R}_{k-m}, \overline{Y}_{k-m}) } } \underbrace{ \left( \prod_{j=k-m}^{k-1}{\color{red} f(Y^{obs}_j | R_j, Y_j) }\right)}_{ {\color{red} f(\overline{Y}_{k}^{obs,m} | \overline{R}_{k}^m, \overline{Y}_{k}^m) }} \underbrace{\left(\prod_{j=k}^{k+m} {\color{red} f(Y^{obs}_j | R_j, Y_j) } \right)}_{ {\color{red} f(\overline{Y}_{k+m+1}^{obs,m} | \overline{R}_{k+m+1}^m, \overline{Y}_{k+m+1}^m) } } \nonumber \\
& {\color{red} f(Y^{obs}_{k+m+1} | R_{k+m+1},Y_{k+m+1}) } \underbrace{ \left(\prod_{j=k+m+2}^{K} {\color{red} f(Y^{obs}_j | R_j, Y_j) }\right) }_{ {\color{red} f(\underline{Y}^{obs}_{k+m+1} | \underline{R}_{k+m+1}, \underline{Y}_{k+m+1}) }}
\nonumber
\end{align}
where the identities shown by the braces follow from the following identifies:
{
\allowdisplaybreaks
\begin{eqnarray*}
f(\overline{Y}_{k-m})  & = &  \prod_{j=1}^{k-m-1} f(Y_j | \overline{Y}_j)  = \prod_{j=1}^{k-m-1} f(Y_j | \overline{Y}^m_j) \\
f(\overline{Y}_{k}^m | \overline{Y}_{k-m})  & = & \prod_{j=k-m}^{k-1} f(Y_j | \overline{Y}_j)  = \prod_{j=k-m}^{k-1} f(Y_j | \overline{Y}^m_j)  \\
f(\underline{Y}_{k-1}^{m+1} | \overline{Y}_{k}^m) & = &  \prod_{j=k}^{k+m}  f(Y_j |  \overline{Y}_{j}^{j-k+m})  = \prod_{j=k}^{k+m}  f(Y_j | \overline{Y}^m_j)    \\
f(Y_{k+m+1}| \underline{Y}_{k-1}^{m+1})  & = &  f(Y_{k+m+1} | \overline{Y}^{m+1}_{k+m+1}) = f(Y_{k+m+1} | \overline{Y}^{m}_{k+m+1}) \\
f(\underline{Y}_{k+m+1} |\underline{Y}_{k-1}^{m+2}  ) & = & \prod_{j=k+m+2}^{K}  f(Y_j |  \overline{Y}_{j}^{j-k})  = \prod_{j=k+m+2}^{K}   f(Y_j | \overline{Y}^m_j)  \\ 
f(\overline{R}_{k-m} | \overline{Y}_{k-m}, \overline{O}_k^m )  & = & \prod_{j=1}^{k-m-1}  f(R_j| \overline{Y}_{k-m}, \underline{R}_j^{k-j-m-1}, \overline{O}_k^m ) \\
& = & \prod_{j=1}^{k-m-1} f(R_j| \overline{Y}_{k-m}, \underline{O}_j^{k-j-1} ) = \prod_{j=1}^{k-m-1} f(R_j| \overline{Y}_{j+1}^{m+1}, \underline{O}_j^{m} )  \\
f(\overline{R}^m_{k} | \overline{Y}_{k}, \underline{O}_{k-1}^{m+1} ) & = & \prod_{j=k-m}^{k-1} f(R_j| \overline{Y}_{k}, \underline{R}_j^{k-j-1},\underline{O}_{k-1}^{m+1}  )  \\
& = & \prod_{j=k-m}^{k-1} f(R_j| \overline{Y}_{k}, \underline{O}_j^{k-j+m} ) = \prod_{j=k-m}^{k-1} f(R_j| \overline{Y}^{m+1}_{j+1}, \underline{O}_j^{m} ) \\
f(\underline{R}_{k-1}^{m+1} | \overline{Y}^{2m+1}_{k+m+1},\underline{O}_{k+m}) & = &  \prod_{j=k}^{k+m} f(R_j| \overline{Y}^{2m+1}_{k+m+1}, \underline{R}_j^{k+m-j},\underline{O}_{k+m})  \\
& = & \prod_{j=k}^{k+m} f(R_j| \overline{Y}^{2m+1}_{k+m+1}, \underline{O}_j) =
\prod_{j=k}^{k+m} f(R_j| \overline{Y}^{m+1}_{j+1}, \underline{O}_j^{m} )\\
f(R_{k+m+1}| \underline{Y}_{k-1}^{m+2},\underline{O}_{k+m+1})  & = & f(R_{k+m+1}| \overline{Y}_{k+m+2}^{m+2},\underline{O}_{k+m+1})  = f(R_{k+m+1}| \overline{Y}_{k+m+2}^{m+1},\underline{O}_{k+m+1}^m)  \\
f(\underline{R}_{k+m+1}| \underline{Y}_{k-1}) & = & \prod_{j=k+m+2}^{K} f(R_j| \underline{Y}_{k-1}, \underline{R}_j)  \\
& = & \prod_{j=k+m+2}^{K} f(R_j| \underline{Y}_{k-1}, \underline{O}_j)  = \prod_{j=k+m+2}^{K} f(R_j| \overline{Y}^{m+1}_{j+1}, \underline{O}_j^m ) \\
{\color{red} f(\overline{Y}^{obs}_{k-m} | \overline{R}_{k-m}, \overline{Y}_{k-m}) } &= & \prod_{j=1}^{k-m-1} {\color{red} f(Y^{obs}_j |\overline{Y}^{obs}_j,\overline{R}_{k-m}, \overline{Y}_{k-m}) } = \prod_{j=1}^{k-m-1} {\color{red} f(Y^{obs}_j | R_j, Y_j) } \\
{\color{red} f(\overline{Y}_{k}^{obs,m} | \overline{R}_{k}^m, \overline{Y}_{k}^m) } & = & \prod_{j=k-m}^{k-1}{\color{red} f(Y^{obs}_j | \overline{Y}_j^{obs,j-k+m},\overline{R}_{k}^m, \overline{Y}_{k}^m) } = \prod_{j=k-m}^{k-1}{\color{red} f(Y^{obs}_j | R_j, Y_j) }  \\
{\color{red} f(\overline{Y}_{k+m+1}^{obs,m} | \overline{R}_{k+m+1}^m, \overline{Y}_{k+m+1}^m) } & = & \prod_{j=k}^{k+m} {\color{red} f(Y^{obs}_j | \overline{Y}_j^{obs,j-k-1},\overline{R}_{k+m+1}^m, \overline{Y}_{k+m+1}^m) }  = \prod_{j=k}^{k+m} {\color{red} f(Y^{obs}_j | R_j, Y^{(1)}_j) }  \\
{\color{red} f(\underline{Y}^{obs}_{k+m+1} | \underline{R}_{k+m+1}, \underline{Y}_{k+m+1}) } & = &  \prod_{j=k+m+2}^{K} {\color{red} f(Y^{obs}_j | \overline{Y}_j^{obs,j-k-m-2}, \underline{R}_{k+m+1}, \overline{Y}_{k+m+1}) }  \\
& = & \prod_{j=k+m+2}^{K} {\color{red} f(Y^{obs}_j | R_j, Y_j) } \end{eqnarray*}}
For all identities, the first equality follows by the chain rule (except the first equalities for terms
$f(Y_{k+m+1}| \underline{Y}_{k-1}^{m+1})$ and
$f(R_{k+m+1}| \underline{Y}_{k-1}^{m+2},\underline{O}_{k+m+1})$,
which follow by definition).  For the first five identities, the second equality follows by application of the model assumptions.  Specifically, the assumptions imply that every variable is independent of non-parental non-descendants given its parents in the DAG corresponding to the factorization (\ref{eqn:dag-f}).  The second equality then follows by noting that the only elements of the conditioning sets for a given variable are variables that are its parents, and possibly other variables prior in the ordering (\ref{eqn:order}), which are always non-descendants.
For the next five identifies, the second equality follows since relevant $Y^{obs}_l$'s can be added to the conditioning sets as they are deterministic functions of $R_l$'s and $Y_l$'s that are already in the conditioning sets, and the third equality follows by the model assumptions using the same arguments discussed for the first five identities.
For last four identities, the second equality follows because the conditioning set for each variable includes the variables which determine it, thus making all additional variables in the conditioning set irrelevant. 

With these identities, (7) can be re-expressed as
\begin{align}
& f(\overline{Y}_{k-m})   f(\overline{Y}_{k}^m | \overline{Y}_{k-m}) f(\underline{Y}_{k-1}^{m+1} | \overline{Y}_{k}^m)  f(Y_{k+m+1}| \underline{Y}_{k-1}^{m+1})  f(\underline{Y}_{k+m+1} |\underline{Y}_{k-1}^{m+2}  )  \nonumber\\
& f(\overline{R}_{k-m} | \overline{Y}_{k-m}, \overline{O}_k^m )  f(\overline{R}^m_{k} | \overline{Y}_{k}, \underline{O}_{k-1}^{m+1} ) f(\underline{R}_{k-1}^{m+1} | \overline{Y}^{2m+1}_{k+m+1},\underline{O}_{k+m}) f(R_{k+m+1}| \underline{Y}_{k-1}^{m+2},\underline{O}_{k+m+1})  \nonumber\\
& f(\underline{R}_{k+m+1}| \underline{Y}_{k-1}) {\color{red} f(\overline{Y}^{obs}_{k-m} | \overline{R}_{k-m}, \overline{Y}_{k-m}) }
{\color{red} f(\overline{Y}_{k}^{obs,m} | \overline{R}_{k}^m, \overline{Y}_{k}^m) } 
{\color{red} f(\overline{Y}_{k+m+1}^{obs,m} | \overline{R}_{k+m+1}^m, \overline{Y}_{k+m+1}^m) }  \nonumber\\
& {\color{red} f(Y^{obs}_{k+m+1} | R_{k+m+1}, Y_{k+m+1}) }
{\color{red} f(\underline{Y}_{k+m+1}^{obs,m} | \underline{R}_{k+m+1}^m, \underline{Y}_{k+m+1}^m) }.
\nonumber
\end{align}
We have thus established that the full data law also factorizes with respect to the ``reduced'' DAG shown in Fig.~\ref{fig:example} (a).

For case (b), the full law (6) can be written as (7) without the first, sixth and eleventh terms.  Using the above identities, the full law can then be written as 
\begin{align}
&  f(\overline{Y}_{k}^m | \overline{Y}_{k-m}) f(\underline{Y}_{k-1}^{m+1} | \overline{Y}_{k}^m)  f(Y^{(1)}_{k+m+1}| \underline{Y}_{k-1}^{m+1}) 
f(\underline{Y}_{k+m+1} |\underline{Y^{(1)}}_{k-1}^{m+2}  )  \nonumber \\
& f(\overline{R}^m_{k} | \overline{Y}_{k}, \underline{O}_{k-1}^{m+1} ) f(\underline{R}_{k-1}^{m+1} | \overline{Y}^{2m+1}_{k+m+1},\underline{O}_{k+m}) f(R_{k+m+1}| \underline{Y}_{k-1}^{m+2},\underline{O}_{k+m+1}) f(\underline{R}_{k+m+1}| \underline{Y}_{k-1})  \nonumber\\
&
{\color{red} f(\overline{Y}_{k}^{obs,m} | \overline{R}_{k}^m, \overline{Y}_{k}^m) } 
{\color{red} f(\overline{Y}_{k+m+1}^{obs,m} | \overline{R}_{k+m+1}^m, \overline{Y}_{k+m+1}^m) } {\color{red} f(Y^{obs}_{k+m+1} | R_{k+m+1}, Y_{k+m+1}) } \nonumber \\
& {\color{red} f(\underline{Y}_{k+m+1}^{obs,m} | \underline{R}_{k+m+1}^m, \underline{Y}_{k+m+1}^m) }.
\nonumber
\end{align}
Here, the full law factorizes with respect to the ``reduced'' DAG shown in Fig.~\ref{fig:example} (b).  

For case (c), the full law (6) can be written as (7) without the fifth, tenth and fifteenth terms.  Using the above identities, the full law can then be written as 
\begin{align}
& f(\overline{Y}_{k-m})   f(\overline{Y}_{k}^m | \overline{Y}_{k-m}) f(\underline{Y}_{k-1}^{m+1} | \overline{Y}_{k}^m)  f(Y_{k+m+1}| \underline{Y}_{k-1}^{m+1})    \nonumber \\
& f(\overline{R}_{k-m} | \overline{Y}_{k-m}, \overline{O}_k^m )  f(\overline{R}^m_{k} | \overline{Y}_{k}, \underline{O}_{k-1}^{m+1} ) f(\underline{R}_{k-1}^{m+1} | \overline{Y}^{2m+1}_{k+m+1},\underline{O}_{k+m}) f(R_{k+m+1}| \underline{Y}_{k-1}^{m+2},\underline{O}_{k+m+1})  \nonumber \\
& {\color{red} f(\overline{Y}^{obs}_{k-m} | \overline{R}_{k-m}, \overline{Y}_{k-m}) }
{\color{red} f(\overline{Y}_{k}^{obs,m} | \overline{R}_{k}^m, \overline{Y}_{k}^m) } {\color{red} f(\overline{Y}_{k+m+1}^{obs,m} | \overline{R}_{k+m+1}^m, \overline{Y}_{k+m+1}^m) } \nonumber \\
& {\color{red} f(Y^{obs}_{k+m+1} | R_{k+m+1}, Y_{k+m+1}) }.
\nonumber
\end{align}
Here, the full law factorizes with respect to the ``reduced'' DAG shown in Fig.~\ref{fig:example} (c).  

Any d-separation statement in Figures.~\ref{fig:example} (a), (b) and (c) implies the corresponding conditional independence statement in the
full data law, as expressed in (8), (9) and (10), respectively.  This immediately implies the conditional independence in the statement of the lemma holds.

\end{proof}

\secondind*
\begin{proof}
This follows from the local Markov property of DAGs, which states that in a graph ${\cal G}$,  $V \ci \nd_{\cal G}(V) \setminus \pa_{\cal G}(V) \mid \pa_{\cal G}(V)$.  Note that $\overline{Y}_{k+1}^{m+1},\underline{O}^{m}_{k}$ are parents of $R_k$ and $O_{k+m+1}$ is a non-parental, non-descendant of $R_k$.
\end{proof}

\begin{figure}
	\begin{center}
		\begin{tikzpicture}[>=stealth, node distance=2.5cm]
		\tikzstyle{format} = [draw, very thick, circle, minimum size=5.0mm,
		inner sep=0pt]
		\tikzstyle{unode} = [draw, very thick, circle, minimum size=1.0mm,
		inner sep=0pt]
		\tikzstyle{square} = [draw, very thick, rectangle, minimum size=4mm]
		\tikzstyle{type1} = [rectangle, rounded corners, minimum width=0.5cm, minimum height=0.1cm,text centered, draw=black, fill=blue!10, text width=1.5cm]
		\tikzstyle{type2} = [rectangle, rounded corners, minimum width=0.5cm, minimum height=0.1cm,text centered, draw=black, fill=green!10, text width=1.5cm]
		\tikzstyle{type3} = [rectangle, rounded corners, minimum width=0.5cm, minimum height=0.1cm,text centered, draw=black, fill=red!10, text width=1.5cm]

		\begin{scope}[xshift=0.0cm]
		\path[->, very thick]
		node[] (x11) {$\overline{Y}_{k-m}$}
		node[type1,right of=x11] (x12) {$\overline{Y}^m_{k}$}
		node[right of=x12] (x13) {$\underline{Y}_{k-1}^{m+1}$}
		node[right of=x13] (x14) {$Y_{k+m+1}$}
		node[right of=x14] (x15) {$\underline{Y}_{k+m+1}$}
		
		node[below of=x11] (r1) {$\overline{R}_{k-m}$}
		node[type2,below of=x12] (r2) {$\overline{R}^m_{k}$}
		node[type1,below of=x13] (r3) {$\underline{R}_{k-1}^{m+1}$}
		node[type3,below of=x14] (r4) {$R_{k+m+1}$}
		node[below of=x15] (r5) {$\underline{R}_{k+m+1}$}
		
		node[below of=r1] (x1) {$\overline{Y}^{obs}_{k-m}$}
		node[below of=r2] (x2) {$\overline{Y}^{obs,m}_{k}$}
		node[type1,below of=r3] (x3) {$\underline{Y}_{k-1}^{obs,m+1}$}
		node[type3,below of=r4] (x4) {$Y^{obs}_{k+m+1}$}
		node[below of=r5] (x5) {$\underline{Y}^{obs}_{k+m+1}$}
		
		(r1) edge[red] (x1)
		(r2) edge[red] (x2)
		(r3) edge[red] (x3)
		(r4) edge[red] (x4)
		(r5) edge[red] (x5)

		(x11) edge[brown] (r1)
		(x12) edge[brown] (r2)
		(x13) edge[brown] (r3)
		(x14) edge[brown] (r4)
		(x15) edge[brown] (r5)

		(x11) edge[red, bend right=40] (x1)
		(x12) edge[red, bend right=40] (x2)
		(x13) edge[red, bend right=40] (x3)
		(x14) edge[red, bend right=40] (x4)
		(x15) edge[red, bend right=40] (x5)
		
		(r2) edge[blue] (r1)
		(r3) edge[blue] (r2)
		(r4) edge[blue] (r3)
		(r5) edge[blue] (r4)

		(x2) edge[blue] (r1)
		(x3) edge[blue] (r2)		
		(x4) edge[blue] (r3)
		(x5) edge[blue] (r4)		
		
		(x11) edge[blue] (r2)
		(x12) edge[blue] (r3)
		(x13) edge[blue] (r4)
		(x14) edge[blue] (r5)
		
		(x14) edge[blue] (x15)
		(x13) edge[blue] (x14) 
		(x12) edge[blue] (x13)
		(x11) edge[blue] (x12)

		(x13) edge[blue, bend right=-15] (x15)
		(r5) edge[blue, bend left=-15] (r3)
		(x13) edge[blue] (r5)
		(x5) edge[blue] (r3)
		
		node[below of=x3, yshift=1.7cm, xshift=0.0cm] (l) {$(a)$}
		;
		\end{scope}

		\begin{scope}[yshift=-7.0cm]
		\path[->, very thick]
		node[] (x11) {}
		node[type1,right of=x11] (x12) {$\overline{Y}^m_{k}$}
		node[right of=x12] (x13) {$\underline{Y}_{k-1}^{m+1}$}
		node[right of=x13] (x14) {$Y_{k+m+1}$}
		node[right of=x14] (x15) {$\underline{Y}_{k+m+1}$}
		
		node[below of=x11] (r1) {}
		node[type2,below of=x12] (r2) {$\overline{R}^m_{k}$}
		node[type1,below of=x13] (r3) {$\underline{R}_{k-1}^{m+1}$}
		node[type3,below of=x14] (r4) {$R_{k+m+1}$}
		node[below of=x15] (r5) {$\underline{R}_{k+m+1}$}
		
		node[below of=r1] (x1) {}
		node[below of=r2] (x2) {$\overline{Y}^{obs,m}_{k}$}
		node[type1,below of=r3] (x3) {$\underline{Y}_{k-1}^{obs,m+1}$}
		node[type3,below of=r4] (x4) {$Y^{obs}_{k+m+1}$}
		node[below of=r5] (x5) {$\underline{Y}^{obs}_{k+m+1}$}
		
		(r2) edge[red] (x2)
		(r3) edge[red] (x3)
		(r4) edge[red] (x4)
		(r5) edge[red] (x5)

		(x12) edge[brown] (r2)
		(x13) edge[brown] (r3)
		(x14) edge[brown] (r4)
		(x15) edge[brown] (r5)

		(x12) edge[red, bend right=40] (x2)
		(x13) edge[red, bend right=40] (x3)
		(x14) edge[red, bend right=40] (x4)
		(x15) edge[red, bend right=40] (x5)
		
		(r3) edge[blue] (r2)
		(r4) edge[blue] (r3)
		(r5) edge[blue] (r4)

		(x3) edge[blue] (r2)		
		(x4) edge[blue] (r3)
		(x5) edge[blue] (r4)		
		
		(x12) edge[blue] (r3)
		(x13) edge[blue] (r4)
		(x14) edge[blue] (r5)
		
		(x14) edge[blue] (x15)
		(x13) edge[blue] (x14) 
		(x12) edge[blue] (x13)

		(x13) edge[blue, bend right=-15] (x15)
		(r5) edge[blue, bend left=-15] (r3)
		(x13) edge[blue] (r5)
		(x5) edge[blue] (r3)
		
		node[below of=x3, yshift=1.7cm, xshift=0.0cm] (l) {$(b)$}
		;
		\end{scope}

		\begin{scope}[yshift=-14.0cm]
		\path[->, very thick]
		node[] (x11) {$\overline{Y}_{k-m}$}
		node[type1,right of=x11] (x12) {$\overline{Y}^m_{k}$}
		node[right of=x12] (x13) {$\underline{Y}_{k-1}^{m+1}$}
		node[right of=x13] (x14) {$Y_{k+m+1}$}
		node[right of=x14] (x15) {}
		
		node[below of=x11] (r1) {$\overline{R}_{k-m}$}
		node[type2,below of=x12] (r2) {$\overline{R}^m_{k}$}
		node[type1,below of=x13] (r3) {$\underline{R}_{k-1}^{m+1}$}
		node[type3,below of=x14] (r4) {$R_{k+m+1}$}
		node[below of=x15] (r5) {}
		
		node[below of=r1] (x1) {$\overline{Y}^{obs}_{k-m}$}
		node[below of=r2] (x2) {$\overline{Y}^{obs,m}_{k}$}
		node[type1,below of=r3] (x3) {$\underline{Y}_{k-1}^{obs,m+1}$}
		node[type3,below of=r4] (x4) {$Y^{obs}_{k+m+1}$}
		node[below of=r5] (x5) {}
		
		(r1) edge[red] (x1)
		(r2) edge[red] (x2)
		(r3) edge[red] (x3)
		(r4) edge[red] (x4)

		(x11) edge[brown] (r1)
		(x12) edge[brown] (r2)
		(x13) edge[brown] (r3)
		(x14) edge[brown] (r4)

		(x11) edge[red, bend right=40] (x1)
		(x12) edge[red, bend right=40] (x2)
		(x13) edge[red, bend right=40] (x3)
		(x14) edge[red, bend right=40] (x4)
		
		(r2) edge[blue] (r1)
		(r3) edge[blue] (r2)
		(r4) edge[blue] (r3)

		(x2) edge[blue] (r1)
		(x3) edge[blue] (r2)		
		(x4) edge[blue] (r3)
		
		(x11) edge[blue] (r2)
		(x12) edge[blue] (r3)
		(x13) edge[blue] (r4)
		
		(x13) edge[blue] (x14) 
		(x12) edge[blue] (x13)
		(x11) edge[blue] (x12)

		
		node[below of=x3, yshift=1.7cm, xshift=0.0cm] (l) {$(c)$}
		;
		\end{scope}

		\end{tikzpicture}
	\end{center}
	\caption{
	Simplified graphical representation of the $m$th-order Markov models, for different values of $k$. (a)  $m < k < K - m - 1$, (b) $k \leq m$, (c) $k = K - m - 1$. Red edges represent deterministic relationships. Brown edges represent edges introduced through non-zero $\alpha$ in exponential tilting for sensitivity analysis. Blue edges indicate a non-deterministic probabilistic dependence of a variable on its parents.
	}
	\label{fig:example}
\end{figure}

\idinduction*

\begin{proof}
The proof proceeds by induction, forward in time.  First, when $k=1$, $f(\overline{Y}_2^{m+1},\underline{O}^{m+1}_1)$ is identified via the following formula:
\begin{eqnarray*}
f(\overline{Y}_2^{m+1},\underline{O}^{m+1}_1) & = & f(Y_1,\underline{O}^{m+1}_1) \\
& = & f(O_{m+2} | Y_1,\underline{O}^{m}_1) f(Y_1 | \underline{O}^{m}_1) f(\underline{O}^{m}_1) \\
& = & f(O_{m+2} | Y_1,R_1=1,\underline{O}^{m}_1) f(Y_1 | \underline{O}^{m}_1) f(\underline{O}^{m}_1) \\
& = & f(O_{m+2}| R_1=1,\underline{O}^{m}_1)  \\
&& \hspace*{0.2in} \left\{ f(Y_1 | R_1 = 1, \underline{O}_1^m )P(R_1=1|\underline{O}^{m}_1) + f(Y_1  | R_1 = 0, \underline{O}_1^m ) P(R_1=0|\underline{O}^{m}_1)  \right\} \\
&& \hspace*{0.2in} f(\underline{O}_1^{m}) \\
& = & f(O_{m+2},Y_1 | R_1=1,\underline{O}^{m}_1)  \\
&& \hspace*{0.2in} \left\{ P(R_1=1|\underline{O}^{m}_1) + P(R_1=0|\underline{O}^{m}_1) \frac{ \exp\{ \alpha_1 Y^{(1)}_1\} }{ c_1(\underline{O}^{m}_1;\alpha_1)} \right\}f(\underline{O}_1^{m}), 
\end{eqnarray*}
where the third equality follows by Lemma 2,  and the fifth equality by Equation \ref{eqn:permutation-model0c}. 
All quantities on the right hand side of last quality are identified from the distribution of the observed data.
 
Suppose that $f(\overline{Y}_{k}^{m+1},\underline{O}^{m+1}_{k-1})$ is identified, where $k\leq K$. We want to show that $f(\overline{Y}_{k+1}^{m+1},\underline{O}^{m+1}_{k})$ is identified. Then,

 \begin{eqnarray}
f(\overline{Y}_{k+1}^{m+1},\underline{O}^{m+1}_{k})& = & f(O_{k+m+1} | \overline{Y}_{k+1}^{m+1},\underline{O}^{m}_k)^{I(k \leq K -m -1)} f(Y_k | \overline{Y}_{k}^{m}, \underline{O}^{m}_k) f(\overline{Y}_{k}^{m},\underline{O}^{m}_k)  \nonumber \\
& = & f(O_{k+m+1} |R_k=1, \overline{Y}_{k+1}^{m+1},\underline{O}^{m}_k)^{I(k \leq K - m - 1)} f(Y_k | \overline{Y}_{k}^{m}, \underline{O}^{m}_k) f(\overline{Y}_{k}^{m},\underline{O}^{m}_k)  \nonumber \\
& = & f(O_{k+m+1} | R_k=1, \overline{Y}_{k+1}^{m+1}, \underline{O}^{m}_k)^{I(k \leq K -m - 1)}  \nonumber \\
&& \Big\{ f(Y_k | R_k = 1, \overline{Y}_k^m, \underline{O}_k^m ) P(R_k=1|\overline{Y}_{k}^{m},\underline{O}^{m}_{k}) + \nonumber\\
&& f(Y_k | R_k = 0, \overline{Y}_k^m, \underline{O}_k^m )P(R_k=0|\overline{Y}_{k}^{m},\underline{O}^{m}_{k}) \Big\} \nonumber\\
&& f(\overline{Y}_{k}^{m},\underline{O}^{m}_{k} ) \nonumber  \\
& = & f(O_{k+m+1}| R_k=1, \overline{Y}_{k+1}^{m+1}, \underline{O}^{m}_k)^{I(k \leq K - m - 1)} f(Y_k \mid R_k = 1 , \overline{Y}_k^m, \underline{O}_k^m) \nonumber \\
&& \left\{ P(R_k=1|\overline{Y}_{k}^{m},\underline{O}^{m}_{k}) + P(R_k=0|\overline{Y}_{k}^{m},\underline{O}^{m}_{k}) \frac{ \exp\{ \alpha_k Y_k \}}{c_k(\overline{Y}_{k}^{m},\underline{O}^{m}_{k};\alpha_k) } \right\}\nonumber\\
&&f(\overline{Y}_{k}^{m},\underline{O}^{m}_{k} ), \label{eqn:induction-last-equality} 
\end{eqnarray}
where the first equality follows by chain rule, the second equality follows by Lemma 2, the third equality by the marginal rule of probability, and the fourth equality follows by (\ref{eqn:permutation-model0c}).

%
%

The right hand side of (\ref{eqn:induction-last-equality}) is a function of $f(\overline{Y}_{k}^{m},\underline{O}^{m+2}_{k-1} )$ through probability operations of marginalization and conditioning.  When $k=1$, this is $f(O_{0}^{m+2})$ which is a function of observed data. To see that this is identified for $k > 1$, it is first useful to note that $ f(\overline{Y}_{k}^{m},\underline{O}^{m+1}_{k-1} )$ is identified by the induction hypothesis $f(\overline{Y}_k^{m+1}, \underline{O}_{k-1}^{m+1})$ since it is equal to $\sum_{Y_{k-m-1}} f(\overline{Y}_{k}^{m+1},\underline{O}^{m+1}_{k-1} )$ if $k>m+1$ and equal to $f(\overline{Y}_{k}^{m+1},\underline{O}^{m+1}_{k-1} )$ if $k \leq m+1$.  Now,  consider the following two cases: 
\begin{enumerate}
\item If $k > K-m-1$, then $ f(\overline{Y}_{k}^{m},\underline{O}^{m+2}_{k-1} )= f(\overline{Y}_{k}^{m},\underline{O}^{m+1}_{k-1} )$, which is identified as discussed above.
 \item If $k \leq K-m-1$, then 
 \begin{eqnarray*}  
f(\overline{Y}_{k}^{m},\underline{O}^{m+2}_{k-1} ) & = & f(O_{k+m+1} | \overline{Y}^m_{k},\underline{O}^{m+1}_{k-1} ) f(\overline{Y}^m_{k},\underline{O}^{m+1}_{k-1}) \\ 
& = &  f(O_{k+m+1} | \overline{Y}^m_{k},\underline{O}^{m+1}_{k-1}, \overline{R}_k^m=1 ) f(\overline{Y}^m_{k},\underline{O}^{m+1}_{k-1}) 
\end{eqnarray*}
where the second equality follows by Lemma 1.  The first term on the right hand side  of the second equality depends on the distribution of the observed data and the second term is identified as discussed above.
\end{enumerate}
\end{proof}

\subsection{Example: $K=7$, $m=2$}

Here, $K-m-1=4$. 
Lemma 1 implies that $R_1$ is independent of $O_5$ given $(Y_1,O_2,O_3,O_4)$, $(R_1,R_2)$ is independent of $O_6$ given $(Y_1,Y_2,O_3,O_4,O_5)$, $(R_2,R_3)$ is independent of $O_7$ given $(Y_2,Y_3,O_4,O_5,O_6)$.

Lemma 2 implies that $R_1$ is independent of $O_4$ given $(Y_1,O_2,O_3)$, $R_2$ is independent of $O_5$ given $(Y_1,Y_2,O_3,O_4)$, $R_3$ is independent of $O_6$ given $(Y_1,Y_2,Y_3,O_4,O_5)$ and $R_4$ is independent of $O_7$ given $(Y_2,Y_3,Y_4,O_5,O_6)$.  

Lemma 3 implies the following, where we highlight pieces of the observed data distribution that need to be estimated in {\color{red} red}:

\begin{itemize}
    \item $k=1$:  $f(Y_1,O_2,O_3,O_4)$ is equal to 
    \begin{align*}
    & f(O_4|R_1=1,Y_1,O_2,O_3) f(Y_1=y_1|R_1=1,O_2,O_3) \times  \\
    & \; \; \; \left\{ P(R_1=1|O_2,O_3) + P(R_1=0|O_2,O_3) \frac{\exp(\alpha_1 Y_1) }{c_1(O_2,O_3;\alpha_1)} \right\} f(O_2,O_3)
    \end{align*}
    which is a function of {\color{red} $f(O_1, O_2, O_3, O_4)$} and thus identified.
    \item $k=2$: $f(Y_1,Y_2,O_3,O_4,O_5)$ is equal to
    \begin{align*}
    & f(O_5|R_2=1,Y_1,Y_2,O_3,O_4) f(Y_2| R_2=1, Y_1,O_3,O_4) \times \\
    & \; \; \; \left\{ P(R_2=1|Y_1,O_3,O_4) + P(R_2=0|Y_1,O_3,O_4) \frac{\exp(\alpha_2 Y_2) }{c_2(Y_1,O_3,O_4;\alpha_2)} \right\} f(Y_1,O_3,O_4)
    \end{align*}
    This is a function of $f(Y_1,O_2,O_3,O_4,O_5)$ which is equal to
    \[
    {\color{red} f(O_5|Y_1,O_2,O_3,O_4,R_1=1)} f(Y_1,O_2,O_3,O_4),
    \]
    where the first term (follows from Lemma 1) is identified and the second term is identified as shown for $k=1$.
    \item $k=3$: $f(Y_1,Y_2,Y_3,O_4,O_5,O_6)$ is equal to
   \begin{align*}
    & f(O_6|R_3=1,Y_1,Y_2,Y_3,O_4,O_5) f(Y_3|R_3=1,Y_1,Y_2,O_4,O_5)\times \\
    & \; \; \; \left\{ P(R_3=1|Y_1,Y_2,O_4,O_5) + P(R_3=0|Y_1,Y_2,O_4,O_5) \frac{\exp(\alpha_3 Y_3) }{c_3(Y_1,Y_2,O_4,O_5;\alpha_3)} \right\} f(Y_1,Y_2,O_4,O_5)
    \end{align*}
    This is a function of $f(Y_1,Y_2,O_3,O_4,O_5,O_6)$ which is equal to
    \[
    {\color{red} f(O_6|Y_1,Y_2,O_3,O_4,O_5,R_1=R_2=1)} f(Y_1,Y_2,O_3,O_4,O_5),
    \]
    where the first term (follows from Lemma 1) is identified and the second term is identified as shown for $k=2$.
    \item $k=4$: $f(Y_2,Y_3,Y_4,O_5,O_6,O_7)$ is equal to
     \begin{align*}
    & f(O_7|R_4=1,Y_2,Y_3,Y_4,O_5,O_6) f(Y_4|R_4=1,Y_2,Y_3,O_5,O_6)  \times \\
    & \; \; \; \left\{ P(R_4=1|Y_2,Y_3,O_5,O_6) + P(R_4=0|Y_2,Y_3,O_5,O_6) \frac{\exp(\alpha_4 Y_4) }{c_4(Y_2,Y_3,O_5,O_6;\alpha_4)} \right\} f(Y_2,Y_3,O_5,O_6)
    \end{align*}
    This is a function of $f(Y_2,Y_3,O_4,O_5,O_6,O_7)$ which is equal to
    \[
    {\color{red} f(O_7|Y_2,Y_3,O_4,O_5,O_6,R_2=R_3=1) } \left\{ \sum_{Y_1} f(Y_1,Y_2,Y_3,O_4,O_5,O_6) \right\},
    \]
    where the first term (follows from Lemma 1) is identified and the second term is identified as shown for $k=3$.
    \item $k=5$: $f(Y_3,Y_4,Y_5,O_6,O_7)$ is equal to
     \begin{align*}
    & f(Y_5|R_5=1,Y_3,Y_4,O_6,O_7) \times \\
    & \; \; \; \left\{ P(R_5=1|Y_3,Y_4,O_6,O_7) + P(R_5=0|Y_3,Y_4,O_6,O_7) \frac{\exp(\alpha_5 Y_5) }{c_5(Y_3,Y_4,O_6,O_7;\alpha_5)} \right\} f(Y_3,Y_4,O_6,O_7)
    \end{align*}
    This is a function of $f(Y_3,Y_4,O_5,O_6,O_7)$ which is equal to
    \[
   \sum_{Y_2} f(Y_2,Y_3,Y_4,O_5,O_6,O_7),
    \]
    which is identified as shown for $k=4$.    
    \item $k=6$: $f(Y_4,Y_5,Y_6,O_7)$ is equal to
     \begin{align*}
    & f(Y_6|R_6=1,Y_4,Y_5,O_7) \times \\
    & \; \; \; \left\{ P(R_6=1|Y_4,Y_5,O_7) + P(R_5=0|Y_4,Y_5,O_7) \frac{\exp(\alpha_6 Y_6) }{c_6(Y_4,Y_5,O_7;\alpha_6)} \right\} f(Y_4,Y_5,O_7)
    \end{align*}
    This is a function of $f(Y_4,Y_5,O_6,O_7)$ which is equal to
    \[
   \sum_{Y_3} f(Y_3,Y_4,Y_5,O_6,O_7),
    \]
    which is identified as shown for $k=5$.    
    \item $k=7$: $f(Y_5,Y_6,Y_7)$ is equal to
     \begin{align*}
    & f(Y_7|R_7=1,Y_5,Y_6) \left\{ P(R_7=1|Y_5,Y_6) + P(R_7=0|Y_5,Y_6) \frac{\exp(\alpha_7 Y_7) }{c_7(Y_5,Y_6;\alpha_7)} \right\} f(Y_5,Y_6)
    \end{align*}
    This is a function of $f(Y_5,Y_6,O_7)$ which is equal to
    \[
   \sum_{Y_4} f(Y_4,Y_5,Y_6,O_7),
    \]
    which is identified as shown for $k=6$.   
    \end{itemize}

Theorem 1 implies that
\begin{align*}
& f(Y_1,Y_2,Y_3,Y_4,Y_5,Y_6,Y_7) \\
& = \sum_{O_2,O_3,O_4} f(Y_1,O_2,O_3,O_4) \times \\
& \; \; \; \frac{\sum_{O_3,O_4,O_5} f(Y_1,Y_2,O_3,O_4,O_5) } {\sum_{Y_2} \sum_{O_3,O_4,O_5} f(Y_1,Y_2,O_3,O_4,O_5)} \times \\
& \; \; \; \frac{\sum_{O_4,O_5,O_6} f(Y_1,Y_2,Y_3,O_4,O_5,O_6) } {\sum_{Y_3} \sum_{O_4,O_5,O_6} f(Y_1,Y_2,Y_3,O_4,O_5,O_6)} \times \\
& \; \; \; \frac{\sum_{O_5,O_6,O_7} f(Y_2,Y_3,Y_4,O_5,O_6,O_7) } {\sum_{Y_4} \sum_{O_5,O_6,O_7} f(Y_2,Y_3,Y_4,O_5,O_6,O_7)} \times \\
& \; \; \; \frac{\sum_{O_6,O_7} f(Y_3,Y_4,Y_5,O_6,O_7) } {\sum_{Y_5} \sum_{O_6,O_7} f(Y_3,Y_4,Y_5,O_6,O_7)} \times \\
& \; \; \; \frac{\sum_{O_7} f(Y_4,Y_5,Y_6,O_7) } {\sum_{Y_6} \sum_{O_7} f(Y_4,Y_5,Y_6,O_7)} \times \\
& \; \; \; \frac{f(Y_5,Y_6,Y_7) } {\sum_{Y_7} f(Y_5,Y_6,Y_7)} 
\end{align*}

\subsection{Simulation Strategy}\label{sec:sim_strategy}

In this section, we discuss how to simulate observed data, assuming that we start with specified  distributions for $P(Y_k=1| \overline{Y}_{k}^{m})$ and $P(R_{k}=1| \overline{Y}_{k}^{m}, \underline{O}_{k}^{m} )$ for $k=1,\ldots,K$.  We need to show that $P(R_k=0 |  \overline{Y}^{m+1}_{k+1}, \underline{O}^m_k)$ can be derived from the specified distributions under Model (4). 

We first show that $f(\overline{Y}_{k+1}^{m+1},\underline{O}^{m+1}_{k})$ and $P(Y_k=1|R_k=1,\underline{O}_k^m, \overline{Y}_k^m)$ are computable for $k=1,\ldots,K$.

\begin{lem}\label{lem:restriction_by_dag}
Suppose $P(Y_k=1| \overline{Y}_{k}^{m})$ and $P(R_{k}=1| \overline{Y}_{k}^{m}, \underline{O}_{k}^{m} )$ are specified for $k=1,\ldots,K$.  Then, $f(\overline{Y}_{k+1}^{m+1},\underline{O}^{m+1}_{k})$ and $P(Y_k=1|R_k=1,\underline{O}_k^m, \overline{Y}_k^m)$ are computable for $k=1,\ldots,K$.
\end{lem}

\begin{proof}
The proof follows by induction in reverse time.  First, we know that $$f(\overline{Y}_{K+1}^{m+1},\underline{O}^{m+1}_{K})=f(\overline{Y}_{K+1}^{m+1}) = \sum_{Y_1} \ldots \sum_{Y_{K-m-1}}  \prod_{k=1}^K f(Y_k \mid \overline{Y}_{k}^{m})$$ which can be computed from the specified distributions. Suppose that $f(\overline{Y}_{k+1}^{m+1},\underline{O}^{m+1}_{k})$ is specified.  We want to show that $f(\overline{Y}_{k}^{m+1},\underline{O}^{m+1}_{k-1})$ can be computed. We can write 
\begin{eqnarray*}
f(\overline{Y}_{k}^{m+1},\underline{O}^{m+1}_{k-1}) & = & f(O_k|\underline{O}^{m}_{k},\overline{Y}_{k}^{m+1} ) f(\underline{O}^{m}_{k}|\overline{Y}_{k}^{m+1}) f(\overline{Y}_{k}^{m+1}) \\
&=& f(O_k|\underline{O}^{m}_{k},\overline{Y}_{k}^{m} ) {\color{red} f(\underline{O}^{m}_{k}|\overline{Y}_{k}^{m}) f(\overline{Y}_{k}^{m+1})}
\end{eqnarray*}
where the first equality follows by the Markov property of the DAG, the first piece in red can be computed from $f(\overline{Y}_{k+1}^{m+1},\underline{O}^{m+1}_{k})$, and the second piece in red can be computed from $\{P(Y_j = 1 \mid \overline{Y}_j^m) \}_{j=1}^K$.  Thus, it remains to show that $f(O_k|\underline{O}^{m}_{k},\overline{Y}_{k}^{m} )$ can be computed. Recall that 
\[f(O_k|\underline{O}^{m}_{k},\overline{Y}_{k}^{m} )= \begin{cases}P(R_k=0|\underline{O}^{m}_{k},\overline{Y}_{k}^{m}), \, \text{when} \, R_k=0; \\ P(Y_k=0|R_k=1,\underline{O}_k^m, \overline{Y}_k^m)P(R_k=1|\underline{O}^{m}_{k},\overline{Y}_{k}^{m}), \, \text{when}\, R_k=1, Y_k=0; \\P(Y_k=1|R_k=1,\underline{O}_k^m, \overline{Y}_k^m)P(R_k=1|\underline{O}^{m}_{k},\overline{Y}_{k}^{m}), \, \text{when} \, R_k=1, Y_k=1.\end{cases}\]  Note that $P(R_k=0|\underline{O}^{m}_{k},\overline{Y}_{k}^{m})$ is specified. We need to show that $P(Y_k=1|R_k=1,\underline{O}_k^m, \overline{Y}_k^m)$ is specified.  Note that
\begin{eqnarray*}
P(Y_k=1|\underline{O}_k^m, \overline{Y}_k^m) & = & P(Y_k=1|R_k=1,\underline{O}_k^m, \overline{Y}_k^m) {\color{red} P(R_k=1|\underline{O}_k^m, \overline{Y}_k^m)} +\\
&& \frac{P(Y_k=1|R_k=1,\underline{O}_k^m, \overline{Y}_k^m) \exp(\alpha_k){\color{red} P(R_k=0|\underline{O}_k^m, \overline{Y}_k^m)}}{P(Y_k=1|R_k=1,\underline{O}_k^m, \overline{Y}_k^m) \exp(\alpha_k)+1-P(Y_k=1|R_k=1,\underline{O}_k^m, \overline{Y}_k^m) }  
\end{eqnarray*}
The left hand side is known from $f(\overline{Y}_{k+1}^{m+1},\underline{O}^{m+1}_{k})$ and the red pieces are specified. Thus, it remains to show that we can uniquely solve this equation for the red pieces.  The following lemma, with $x=P(Y_k=1|R_k=1,\underline{O}_k^m, \overline{Y}_k^m)$, $a =  P(Y_k = 1 \mid \underline{O}^m_k, \overline{Y}^m_k)$, $b = P(R_k =1 \mid \underline{O}^m_k, \overline{Y}^m_k)$, $\beta = \exp(\alpha_k)$, establishes that a unique solution in the interval $(0,1)$ exists.  This completes the proof.
\end{proof}

\begin{lem}\label{lem:solve_for_root}

    Let $0 < a < 1$, $0 < b < 1$ and $\beta >0$. Then, 
\[
a = x b + \frac{x \beta}{x \beta + 1- x} (1-b)
\]
has a unique solution in $(0, 1)$.
\end{lem}
\begin{proof}
\begin{align*}
& a = x b + \frac{x \beta}{x \beta + 1- x} (1-b) \\
\implies & a (x \beta + 1-x) = ( x b)(x \beta + 1-x) + x \beta (1-b) \\
\implies & -b(\beta-1) x^2 + (a (\beta-1) - b - (1-b)\beta) x + a = 0\\
\implies & -b(\beta-1) \frac{w^2}{(1+w)^2} + (a (\beta-1) - b - (1-b)\beta) \frac{w}{1+w} + a = 0; \mbox{ where } x=\frac{w}{w+1}\\
\implies & -b(\beta-1) w^2 + (a (\beta-1) - b - (1-b)\beta) w (1+w)  + a (1+w)^2 = 0\\
\implies & (-b(\beta-1) + a (\beta-1) - b - (1-b)\beta + a) w^2 + (a (\beta-1) - b - (1-b)\beta + 2a) w + a = 0\\
\implies & (a-1) \beta w^2 + (a \beta - b - (1-b)\beta + a) w + a = 0
\end{align*}
The last equation is quadratic in $w$. Since $a, b> 0$, then $(a-1)\beta < 0$, $a>0$.  Hence, there is one sign change.  By Descartes' rule of signs, this means there will be a single positive real root for $w$. Then, because $x = \frac{w}{w+1}$, this means that $x$ has a unique solution in $(0, 1)$
\end{proof}

Now, we show that $P(R_k=0 |  \overline{Y}^{m+1}_{k+1}, \underline{O}^m_k)$ can be derived from the specified distributions under Model (4).

\begin{lem}\label{lem:restriction_by_dag1}
Suppose $P(Y_k=1| \overline{Y}_{k}^{m})$ and $P(R_{k}=1| \overline{Y}_{k}^{m}, \underline{O}_{k}^{m} )$ are specified for $k=1,\ldots,K$.  Then, 
$P(R_k=0 |  \overline{Y}^{m+1}_{k+1}, \underline{O}^m_k)$ are computable for $k=1,\ldots,K$. 
\end{lem}

\begin{proof}
We apply Lemma \ref{lem:restriction_by_dag} to obtain $P(Y_k =1 \mid R_k =1, \underline{O}^m_k, \overline{Y}^m_k)$ for all $k =1, \ldots, K$. This ensures that $c_k (\overline{Y}^m_k, \underline{O}^m_k; \alpha_k)$ from Equation (\ref{eqn:permutation-c-factor}) are computable for $k=1,\ldots,K$. This implies that $h_k(\overline{Y}_k^m,\underline{O}_k^m;\alpha_k)$ in Equation (4) are computable for $k=1,\ldots,K$ as they depend on $c_k (\overline{Y}^m_k, \underline{O}^m_k; \alpha_k)$ and $P(R_k=0 \mid \overline{Y}^{m}_{k}, \underline{O}^m_k)$.  Then, $P(R_k=0 |  \overline{Y}^{m+1}_{k+1}, \underline{O}^m_k)$ are computable using Equation (4) for $k=1,\ldots,K$.
\end{proof}

Given $P(Y_k=1| \overline{Y}_{k}^{m})$ and $P(R_{k}=1| \overline{Y}_{k}^{m}, \underline{O}_{k}^{m} )$ and $\alpha_k$ in Equation (4) for $k=1,\ldots,K$, we can simulate observed data as follows: 
\begin{itemize}
    \item Draw $Y_1,\ldots,Y_K$ using Assumption (2) and $P(Y_k=1| \overline{Y}_{k}^{m})$ for $k=1,\ldots,K$.
    \item Draw $R_K$ using $P(R_K=0 |  \overline{Y}^{m+1}_{K+1}, \underline{O}^m_K) = P(R_K=0|Y_{K-m},\ldots,Y_K)$ (computable by Lemma \ref{lem:restriction_by_dag1}). If $R_K=0$, let $Y_K^{obs}=?$.  If $R_K=1$, let $Y_K^{obs}=Y_K$. Let $O_K = (R_K,Y_K^{obs})$.
    \item For $k=K-1,\ldots,1$,
    \begin{itemize}
       \item Draw $R_k$ using $P(R_k=0 |  \overline{Y}^{m+1}_{k+1}, \underline{O}^m_k)$ (computable by Lemma \ref{lem:restriction_by_dag1}). If $R_k=0$, let $Y_k^{obs}=?$.  If $R_k=1$, let $Y_k^{obs}=Y_k$. Let $O_k = (R_k,Y_k^{obs})$.
    \end{itemize}
\end{itemize}

\end{document}